\documentclass[twocolumn]{aastex61}

\newcommand\aastex{AAS\TeX}

\received{}
\revised{}
\accepted{}

\shorttitle{\aastex\ Evolutionary models of red supergiants}
\shortauthors{Chun et al.}

\begin{document}

\title{Evolutionary models of red supergiants: Evidence for a
metallicity-dependent mixing length and implications for Type IIP supernova progenitors}

\correspondingauthor{Sung-Chul Yoon and Sang-Hyun Chun}
\email{yoon@astro.snu.ac.kr}\email{shyunc@astro.snu.ac.kr}
\author{Sang-Hyun Chun}
\affiliation{Department of Physics and Astronomy, Seoul National University, 08826, Seoul, South Korea}

\author{Sung-Chul Yoon}
\affiliation{Department of Physics and Astronomy, Seoul National University, 08826, Seoul, South Korea}
\affiliation{Monash Centre for Astrophysics, School of Physics and Astronomy, Monash University, Victoria 3800, Australia}

\author{Moo-Keon Jung}
\affiliation{Department of Physics, Sogang University, 04107, Seoul, South Korea}

\author{Dong Uk Kim}
\affiliation{Department of Physics, Korea Advanced Institute of Science and Technology, 34141, Daejeon, South Korea}

\author{Jihoon Kim}
\affiliation{Department of Physics and Astronomy, Seoul National University, 08826, Seoul, South Korea}





\begin{abstract}
Recent studies on the temperatures of red supergiants (RSGs) in the local
universe provide us with an excellent observational constraint on RSG models.  
We calibrate the mixing length parameter by
comparing model predictions with the empirical RSG temperatures in
Small and Large Magellanic Clouds, Milky Way, and M31, which are 
inferred from the TiO band and the spectral energy
distribution (SED). Although our RSG models are computed with the MESA code, our
result may be applied to other
 stellar evolution codes, including the BEC and TWIN codes. 
We find evidence that the mixing length increases with
increasing metallicity for both cases where the TiO and SED temperatures of RSGs
are used for the calibration. 
Together with the recent finding of a
similar correlation in low-mass red giants by Tayar et al, this  implies that
the metallicity dependence of the mixing length is a universal feature in
post-main sequence stars of both low and high masses.  Our result implies that
typical Type IIP supernova (SN IIP) progenitors with initial masses of $\sim 10 -
16~M_\odot$ have a radius range of $400 R_\odot \lesssim R \lesssim 800
R_\odot$ regardless of metallicity. As an auxiliary result of this study, 
we find that the hydrogen-rich envelope
mass of SN IIP progenitors for a given initial mass is predicted to be largely
independent of metallicity if the Ledoux criterion with slow semiconvection is
adopted, while the Schwarzschild models predict systematically more massive
hydrogen-rich envelopes for lower metallicity.  
\end{abstract}




\section{Introduction}\label{sec:intro}
The evolution of massive stars in numerical simulations   depends on many
different physical parameters that are related to the efficiency of convective
energy transport, convective overshoot, semi-convection, rotation, mass loss,
binarity, and metallicity, among others~\citep[e.g.,][]{Maeder2000,
Langer2012, Smith2014}.  Given this complexity, details of massive star evolution are
still much debated.  However, there is a solid consensus that massive stars of
B and O types in the mass range of $\sim$9 to $\sim$30~$M_\odot$ become red supergiants (RSGs)
during the post-main sequence phase. 
Most of them would also die as RSGs, unless they underwent
binary interactions during the course of their
evolution~\citep[e.g.,][]{Podsiadlowski1992, Eldridge2013, Yoon2017} and/or
strong enhancement of mass loss during the final evolutionary
stages~\citep[e.g.,][]{Yoon2010, Georgy2012, Meynet2015}.


RSG temperatures are mainly determined by the well-defined Hayashi limit, which
is a sensitive function of the efficiency of convective energy transport and
opacity~\citep{Hayashi1961}. This means that RSG stars can be used as reference
standards for the calibration of some uncertain physical parameters used in
stellar evolution models. In particular, RSG temperatures can provide an
excellent observational constraint on the efficiency of the convective energy
transport in RSG envelopes, which is commonly parameterized by the so-called
mixing length in stellar evolution models~\citep[e.g.,][]{Kipp1990}. 

This approach has become promising over the past decade with observational
studies on RSG temperatures in different environments.  RSG temperatures
have been typically inferred by the model fitting to the TiO absorption band in
optical spectra~\citep[e.g.,][]{Levesque2005, Levesque2006, Massey2009}. The
temperatures from this method show a correlation with metallicity: higher
temperatures at lower metallicities, which is consistent with the prediction
from current several evolutionary models.  Recently, however,
\citet{Davies2013} recalculated the surface temperatures of RSGs in the
Magellanic Clouds using the spectral energy distributions (SEDs) of RSGs.  They
found that the temperatures inferred from the strengths of TiO lines are
systematically lower than those inferred from the SEDs.  Interestingly, no
clear metallicity dependence of RSG temperatures is found with this new
approach~\citep{Davies2015, Gazak2015, Patrick2015}, in contrast to the
conclusions of previous observational studies with the TiO band.  This calls for
us to systematically investigate the metallicity dependence of RSG properties~\citep[cf.][]{Elias1985}. 

In the above-mentioned observational studies, RSG locations on the
Hertzsprung-Russel (RS) diagram have been compared mostly with the Geneva group
models~\citep[e.g.,][]{Ekstrom2012, Georgy2013}.  The Geneva group uses
$\alpha = 1.6H_P$, where $H_P$ is the local pressure scale height at the outer
boundary of the convective core.  Although their models match fairly well the
observed positions of red giants and supergiants of the Milky Way in the HR
diagram,  this value was chosen based on the solar calibration. In this study
we aim to calibrate the mixing-length parameter by comparing the most recent
observations of RSGs with stellar evolution models at various metallicities.  

We choose the MESA code for the model calculations~\citep{Paxton2011,
Paxton2013, Paxton2015}.  MESA is an open source code and currently widely used
for various studies on stellar physics and stellar populations. Our new grid of
models presented in this study will serve as a useful reference for future
studies on massive stars using MESA in the community.  A few studies indicate
that different stellar evolution codes lead to diverse RSG structures for a
given initial condition~\citep{Martins2013, Jones2015}. However, this is mainly
because of different input physics and the Hayashi line that determines RSG
temperatures does not appear to significantly depend on different numerical
codes as long as the same set of stellar structure equations are employed, as
discussed below (Section~\ref{sec:effects}).  Therefore, our calibration of the
mixing length parameter for different metallicities would be of interests to
the users of several other stellar evolution codes as well. 

Using our new grid of models, we also investigate the structure of Type IIP
supernova (SN IIP) progenitors. In particular, both theoretical models on SN
light curves and recent early-time observations of SN IIP imply much smaller
radii of RSGs than predicted by conventional stellar evolution
models~\citep[e.g.,][]{Dessart2013, Gonzalez2015}. In this study we discuss if
the observed RSG temperatures can be consistent with the radii of SN IIP
progenitors inferred from SN studies. 

This paper is organized as follows. In Section~\ref{sec:method}, we describe
the numerical method and physical assumptions adopted for this study.  In
Section~\ref{sec:effects}, we discuss code dependencies of RSG models and the
effects of different physical parameters on the evolution of RSGs.  In
Section~\ref{sec:comp}, we confront our RSG models with observations, and
discuss the metallicity dependence of the convective mixing length.  In
Section~\ref{sec:final} we discuss the implications of our result for  the
final structure of RSGs as SN IIP progenitors. We conclude this work in
Section~\ref{sec:conclusions}.

\section{Numerical methods and physical assumptions} \label{sec:method}

We calculate our models with the MESA code. 
We construct the evolutionary models with both the Schwarzschild and Ledoux
criteria for convection.  With the Ledoux criterion, we consider inefficient
semiconvection with an efficiency parameter of $\alpha_\mathrm{SEM} = 0.01$.
Slow semiconvection is implied by numerical simulations~\citep{Zaussinger2013}.
Note also that extremely fast semiconvection leads to results comparable to
those with the Schwarzschild criterion. Therefore our models can roughly
provide the boundary conditions for inefficient and efficient chemical mixing
in chemically stratified layers.

In the MESA version we use (MESA-8845), the convective region
is determined by the sign change of the difference between the actual
temperature gradient and the adiabatic temperature gradient (the Schwarzschild
criterion) or between the actual temperature gradient and the adiabatic
temperature gradient plus the chemical composition gradient (the Ledoux
criterion; \citealt{Paxton2011}). Recently \citet{Gabriel2014} point
out that such a simple approach may lead to a physically incorrect determination 
of the convective boundary, especially when the chemical composition is discontinuous across the boundary. 
Soon after the completion of the present study, 
a new version of the MESA code has been released where a numerical scheme to rectify 
this issue is introduced~\citep{Paxton2017}. 
As discussed below in Section~\ref{sec:effects}, however, this 
does not significantly affect RSG temperatures.

Convective overshooting is considered with a step function and applied only for
the hydrogen burning core. We calculate model sequences with three different
overshooting parameters: $f_\mathrm{ov} =$~0.05, 0.15, and 0.30, which are given
in units of the local pressure scale height at the upper boundary of the convective
core. Note that~\citet{Martins2013} recently suggested using $f_\mathrm{ov} =
0.1 - 0.2$ based on the distribution of the main sequence stars on the HR
diagram. Rotation is not considered in this study. The hydrogen burning core
tends to be bigger with rapid rotation~\citep{Maeder2000, Heger2000}, and this
effect of rotation on the convective core size can be roughly considered with
different overshooting parameters we use here.  Note also that not a small
fraction of massive stars are slow rotators~\citep[e.g.,][]{Mokiem2006,
Agudelo2013, Simon2014}, in which case the effect of rotation would not be
important. 

For the calibration of the mixing length parameter, we construct RSG models with
four different values: $\alpha =$~1.5, 2.0, 2.5 and 3.0, which are given in
units of the local pressure scale height.  As shown below, the predicted RSG
temperatures with these values can fully cover the observed RSG temperature range.  

We consider four different initial metallicities: $Z=0.004, 0.007, 0.02$, and
$0.04$ scaled with the chemical composition of \citet{Grevesse1998},
which roughly represent the metallicities of Small Magellanic Cloud (SMC),
Large Magellanic Cloud (LMC), our galaxy (Milky Way), and M31, respectively.
Many recent stellar evolution models adopt $Z=0.014$ with the chemical
composition of \citet{Asplund2005} and \citet{Asplund2009} for the Milky Way
metallicity instead of $Z=0.02$ with the composition of Grevesse \& Sauval, but
the predicted RSG temperatures are not significantly affected by the choice
between the two options (see Section~\ref{sec:effects} below). We use the
Dutch scheme for stellar wind mass-loss rates: the mass-loss rate prescriptions
by \citet{Vink2001} for hot stars ($T_\mathrm{eff} > 12500~$K) and
by~\citet{deJager1988} for cool stars($T_\mathrm{eff}  < 12500~$K).  We use the
simple photosphere boundary condition, which means that the full set of the
stellar structure equations are solved up to the outer boundary that is defined
by an optical depth of $\tau = 2/3$.  For other physical parameters including
the opacity table, we employ the default options of the MESA code for massive
stars (i.e., the options in the file 'inlist\_massive\_defaults').  For each
set of physical parameters, we calculate RSG models for different initial
masses in the range from $9M_{\sun}$ to $39M_{\sun}$ with a $2M_\sun$
increment.  All the calculations are stopped when the central temperature
reaches $10^9$~K, from which the envelope structure does not change
significantly until core collapse~\citep[e.g.,][]{Yoon2017}, except for
some 9 and 11~$M_\odot$ models that are stopped at the end of core helium
burning due to a convergence problem.

For the discussion of code dependencies of RSG models in the
section~\ref{sec:effects}, we also present several RSG models with the BEC and
TWIN codes.  The BEC code, which is also often referred to as the STERN code in
the literature~\citep{Heger2000}, has been widely used for the evolutionary
models of massive stars, including the recent Bonn stellar evolution
grids~\citep{Brott2011}.  Like in our MESA models, the convective overshooting is
treated with a step function in the BEC code and the overshooting parameter $f_\mathrm{ov}$
has the same meaning in both cases.  The TWIN code is developed by P. Egglenton
and his collaborators~\citep{Eggleton1971, Eggleton2002, Izzard2006}, and the
WTTS package by \citet{Izzard2006}  has been used for calculating RSG models
with the TWIN code in this study.  In the TWIN code, convective overshooting is
considered by modifying the convection criterion with an overshooting parameter
$\delta_\mathrm{ov}$ as $\nabla_\mathrm{rad} > \nabla_\mathrm{ad} -
\delta_\mathrm{ov}$, where $\nabla_\mathrm{rad}$ and $\nabla_\mathrm{ad}$ are
radiative and adiabatic temperature gradients with respect to pressure in a
logarithmic scale, respectively.  The physical parameters adopted in the BEC
and TWIN codes are described where appropriate in the following section.  Note
also that in all of these codes (MESA, BEC, and TWIN), the optical depth at the
outer boundary is set to be 2/3. 

\section{Effects of physical parameters and code dependencies} \label{sec:effects}

\begin{figure}
\centering
\includegraphics[width=0.9\columnwidth]{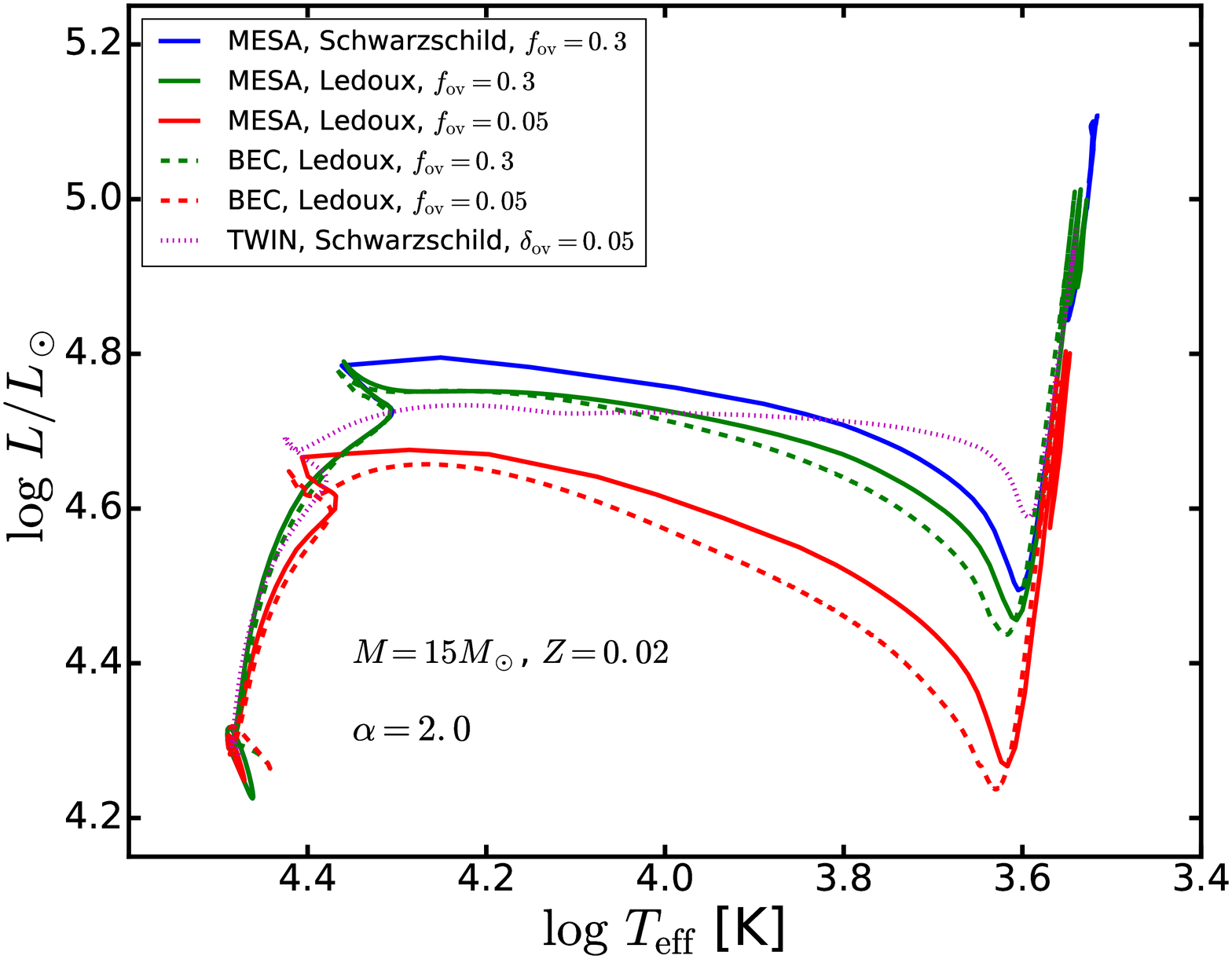}
\includegraphics[width=0.9\columnwidth]{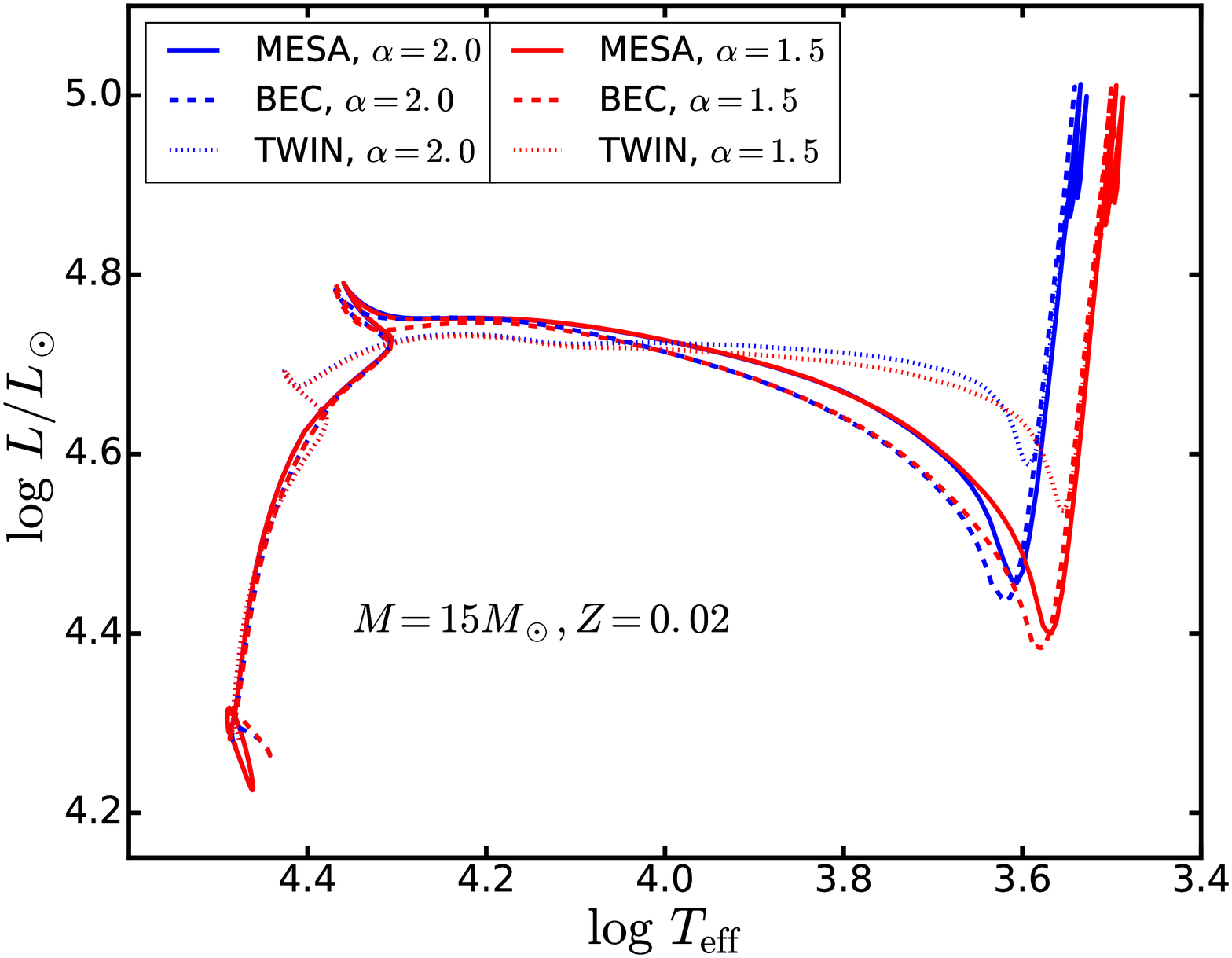}
\caption{\emph{Upper panel:} Comparison of the evolutionary tracks of 15~$M_\odot$
stars at solar metallicity with a mixing length parameter of $\alpha = 2.0$ on
the HR diagram for various overshooting parameters ($f_\mathrm{ov}=0.05$ and
$f_\mathrm{ov}=0.3$ with the MESA and BEC codes, and $\delta_\mathrm{ov} =
0.05$ with the TWIN code; see the text for the details) and convection criteria
(Schwarzschild and Ledoux with a semiconvection parameter of $\alpha_\mathrm{SEM} = 0.01$) calculated with the MESA (solid
lines), BEC (dashed lines), and TWIN (dotted line) codes. 
\emph{Lower panel:} Evolutionary tracks of 15~$M_\odot$ stars at solar metallicity with   
mixing length parameters of $\alpha = 2.0$ (blue)
and $\alpha=1.5$ (red) obtained with the MESA (solid line), BEC (dashed line), and TWIN (dotted line) codes.  The adopted overshooting
parameters are $f_\mathrm{ov} = 0.3$ for the MESA and BEC codes and $\delta = 0.05$ for the TWIN codes. 
The Ledoux criterion for convection with a semiconvection parameter of $\alpha_\mathrm{SEM}=0.01$ is adopted for the MESA and BEC models.  
}  
\label{fig:code}
\end{figure}

\begin{figure}
\centering
\includegraphics[width=0.9\columnwidth]{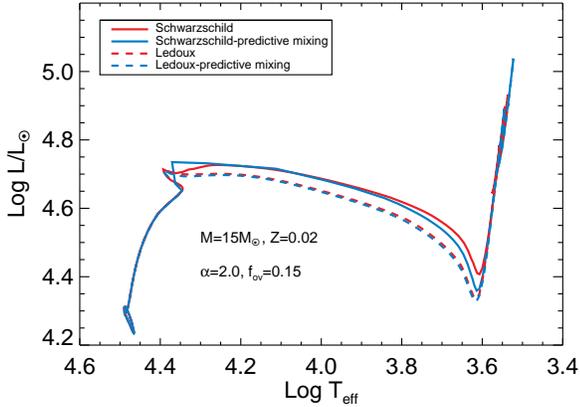}
\caption{Evolutionary tracks of 15~$M_\odot$ stars at $Z=0.02$ with (blue) and without (red) the predictive mixing scheme. 
The adopted mixing length and the overshooting parameter are $\alpha = 2.0$ and $f_\mathrm{ov}=0.15$. 
The solid and dashed lines denote the Schwarzschild and the Ledoux models, respectively.
}\label{fig:predictive}
\end{figure}

\begin{figure}
\centering
\includegraphics[width=0.9\columnwidth]{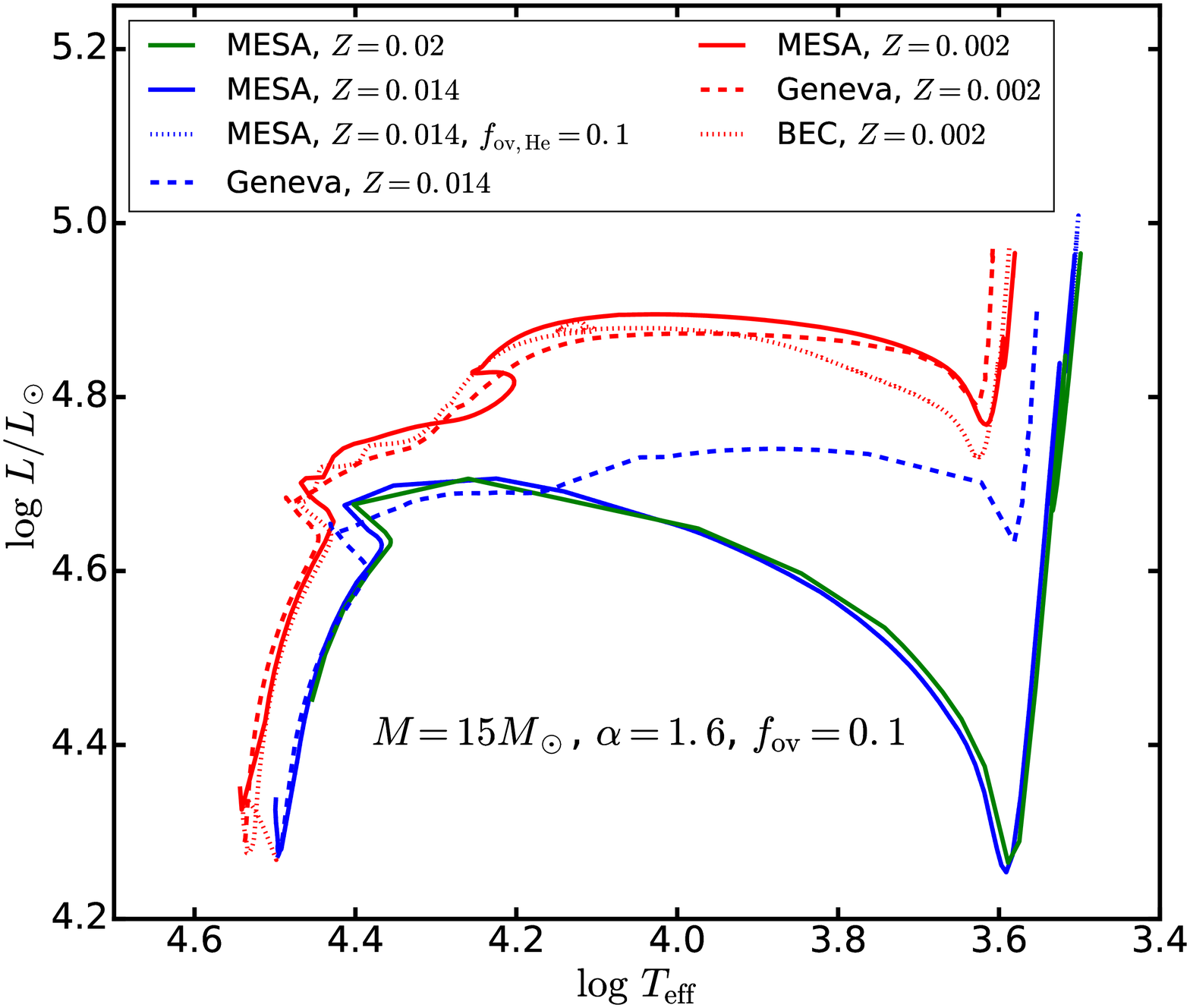}
\caption{Evolutionary tracks of 15~$M_\odot$ stars at $Z=0.02$ with the chemical composition of \citet{Grevesse1998} (green), $Z=0.014$ (blue) with the chemical
composition of \citet{Asplund2009}, and $Z=0.002$ (red) with the \textbf{mixing} length parameter of $\alpha =1.6$
and the overshooting parameter of $f_\mathrm{ov} = 0.1$, calculated with the MESA (solid line), Geneva (dashed line), and BEC (dotted line) codes.
The Schwarzschild criterion is used in all these calculations. 
The blue dotted line, which is overlapped with the blue solid line for the most part, is the $Z=0.014$ MESA model where overshooting is also applied
to the helium-burning core with $f_\mathrm{ov,He} = 0.1$.
}\label{fig:geneva}
\end{figure}

\subsection{Effects of the convection criterion and overshooting}

In Fig.~\ref{fig:code}, we present evolutionary tracks of 15~$M_\odot$ stars on
the HR diagram calculated with MESA, BEC, and TWIN codes using different
convective overshooting parameters and convection criteria. It is shown that
the evolutionary tracks are significantly affected by the adopted physical
parameters.  As already found by numerous studies~\citep[e.g.,][]{Brott2011,
Martins2013}, furthermore, the star have a larger hydrogen burning core with a
larger overshooting parameter. The main sequence width is enlarged and stars
become more luminous with a larger overshooting parameter accordingly.
For a given overshooting parameter in our MESA model,  the evolutionary
tracks of the Schwarzschild and the Ledoux models are identical for most of the
main sequence phase.  This is because the convective core size continues to
decrease on the main sequence, for which both the Schwarzschild and the Ledoux
criteria give the same convective core size.  (see the green and blue lines in
upper panel of Figure~\ref{fig:code}). Readers are referred to
\citet{Martins2013} for more detailed discussion on the effect of various
physical parameters on massive star evolution, in particular on the main
sequence. 

However, once the star reaches the RSG phase, the temperatures obtained with
different overshooting parameters and different convection criteria converge to
almost the same location on the HR diagram, for a given mixing length.  Models
with a larger overshooting parameter tend to have somewhat lower RSG
temperatures as discussed in Section~\ref{sec:comp} in detail, but its effect
is small compared to those of the mixing length and metallicity.  It is shown
that a larger mixing length leads to higher RSG temperatures,  in good
agreement with previous studies~\citep[e.g.][]{Schaller1992}.  

As explained above in Section~\ref{sec:method}, 
a new scheme to determine the boundaries of convective zones
has been introduced in the latest version of MESA~\citep{Paxton2017}. 
We compare the results with and without this so-called predictive mixing scheme 
for a 15~$M_\odot$ star at $Z=0.02$ in  Figure~\ref{fig:predictive}. 
We find that RSG temperatures are not significantly affected by this new scheme and
the main conclusions of our work would not change either. 
However, we note that this new scheme can have important consequences in the inner structures if the Ledoux criterion is used, 
while  models with the Schwarzschild criterion are hardly affected by this. 
In particular, the helium core splitting that has been commonly 
found in previous massive star models using the Ledoux criterion
does not occur with the predictive mixing scheme. 
As a result, the size of the carbon core 
becomes much larger. For example, the helium ($M_\mathrm{HeC}$) and carbon core ($M_\mathrm{CC}$)
masses in a 15~$M_\odot$ star with the Ledoux criterion at the pre-supernova
stage are predicted to be $M_\mathrm{HeC} = 5.0 M_\odot$ and $M_\mathrm{CC} = 3.0~M_\odot$
when the predictive mixing scheme is used, compared to $M_\mathrm{HeC} = 4.9 M_\odot$ and $M_\mathrm{CC} = 1.9~M_\odot$
resulting from the previous method to determine the convective boundaries. 
A detailed investigation of this effect on the pre-supernova structure of massive stars  
would be an interesting subject of future studies. 

\subsection{Effects of metallicity and chemical composition}

The metallicity effect is shown in Figure~\ref{fig:geneva}.  As expected from
the theory of the Hayashi limit~\citep{Hayashi1961}, lower metallicity (hence
lower opacity) leads to higher RSG temperatures for a given $\alpha$.  This
confirms the well known fact that the convective energy transport efficiency
and the opacity are the primary factors that determines RSG temperatures.

In the present study we take the traditional value of $Z=0.02$ with the
chemical composition of ~\citet{Grevesse1998} as the metallicity of Milky Way. Many
recent stellar evolution studies instead assume $Z=0.014$ with the chemical composition
given by \citet{Asplund2005} or \citet{Asplund2009}.  We also compare the two
cases in Figure~\ref{fig:geneva} but the difference in the RSG temperature is
less than 50~K. We conclude that our results do not significantly depend on 
the choice between the two options for the Milky Way metallicity.

\subsection{Code dependencies}

We find some significant dependencies of the adopted numerical codes
\citep[cf.,][]{Martins2013, Jones2015}.   For example, the luminosity with the
Ledoux criterion for a given overshooting parameter during the transition phase
from the end of core hydrogen exhaustion until the beginning of the RSG phase
is somewhat higher with MESA than with BEC (Figure~\ref{fig:code}).  The reason
for this discrepancy despite the same adopted convection parameters is very
difficult to understand, and its clarification is beyond the scope of this
paper.  However, models from MESA, BEC and TWIN codes have very similar RSG
temperatures.  The difference in RSG temperatures along the Hayashi line
resulting from different codes is smaller than $\pm$100~K for a given
luminosity, as long as the same mixing length is adopted.  

On the other hand, the models of the Geneva group, which have been most widely
used in the literature for the comparison with observed RSGs, give
significantly higher RSG temperatures on average, compared to those given by
MESA, BEC and TWIN models.  As an example,  MESA and Geneva tracks for a
15~$\mathrm{M_\odot}$ star with $Z = 0.014$ and $0.002$ are compared in
Fig.~\ref{fig:geneva}, for which the same overshooting parameter
($f_\mathrm{ov}=0.1$) and convection criterion (Schwarzschild) have been
adopted. The Geneva group also considers overshooting above the
helium-burning core, while it is applied only for the hydrogen-burning core in
our MESA models.  Therefore, in the figure, we also include an evolutionary
track for which overshooting of $f_\mathrm{ov,He} = 0.1$ is applied to the
helium-burning core for comparison. 

The Geneva group
adopts the chemical composition of \citet{Asplund2005} except for $^{20}\mathrm{Ne}$ for
which they take the value of \citet{Cunha2006}.  Our MESA models with
$Z=0.014$ adopt the composition of \citet{Asplund2009}, which is essentially
the same with the Geneva model composition.

As shown in the figure, the difference in RSG temperatures along the Hayashi line between GENEVA and
MESA models amounts to about 240~K, which is much bigger than the
difference between MESA and BEC/TWIN models ($ <  100~\mathrm{K}$).
The overshooting  above the helium-burning core does not
change the evolution on the HR diagram except that the luminosity during the last stage 
increases with the overshooting of the helium-burning core.


The reason that the Geneva group code gives distinctively different
results compared to the other cases is difficult to understand. 
One possible reason would be the different numerical schemes adopted
for the outermost layers of the star. 
With the default
atmosphere boundary condition (i.e., the `simple photosphere' option) in MESA,
all the stellar structure equations are fully solved up to the outer boundary,
which is also the case for the BEC and TWIN codes.  The Geneva group solves the
full set of stellar structure equations only for the inner layers, and treats
the envelope and the atmosphere with a reduced set of
equations~\citep{Meynet1997}.  In particular, the energy conservation equation
is not considered in the envelope in the Geneva code.  
Although the luminosity due to the gravitational energy in the RSG envelope
is negligibly small compared to the total luminosity, we still have to investigate 
how the omission of the energy equation can non-linearly influences the envelope structure.  
The treatment of the
step overshooting in the Geneva code is also somewhat different from that of MESA:
the Geneva code uses the adiabatic temperature gradient in the overshooting region instead of
the radiative temperature gradient.  However, this different prescription of
overshooting is not likely to be responsible for such a big temperature
difference of $~ 200$~K, given that the RSG temperatures from the TWIN
code, which also adopts an overshooting scheme different from that of MESA and
BEC as explained above, are very similar to the predictions of the MESA and BEC
models.
 
We tentatively conclude that code dependencies of RSG temperatures along the
Hayashi line are weak as long as the full set of stellar structure equations
are solved up to the outer boundary of the star with a similar boundary
condition.  Our calibration of the mixing length with RSGs in this study can be
relevant not only to MESA but also to several other stellar evolution codes
including BEC and TWIN. A more detailed investigation of the effects of the
numerical schemes on the Hayashi limit is needed for further clarification of
this issue, which we leave as future work.  

\subsection{Additional remarks}

It should also be noted that not all observed RSGs would be on the well defined
Hayashi line. Some of them would be on the way to the Hayashi limit from the
main sequence or the blue loop transition, and some others would be moving away
from it due to the blue loop evolution or strong mass loss.  Therefore, the
observed distribution of RSG temperatures would depend not only on the Hayashi
limit for a given mixing length and metallicity, but also on which fraction of
the RSG lifetime is spent for the transition phases to/from the Hayashi limit.

\begin{figure}
\centering
\includegraphics[width=0.9\columnwidth]{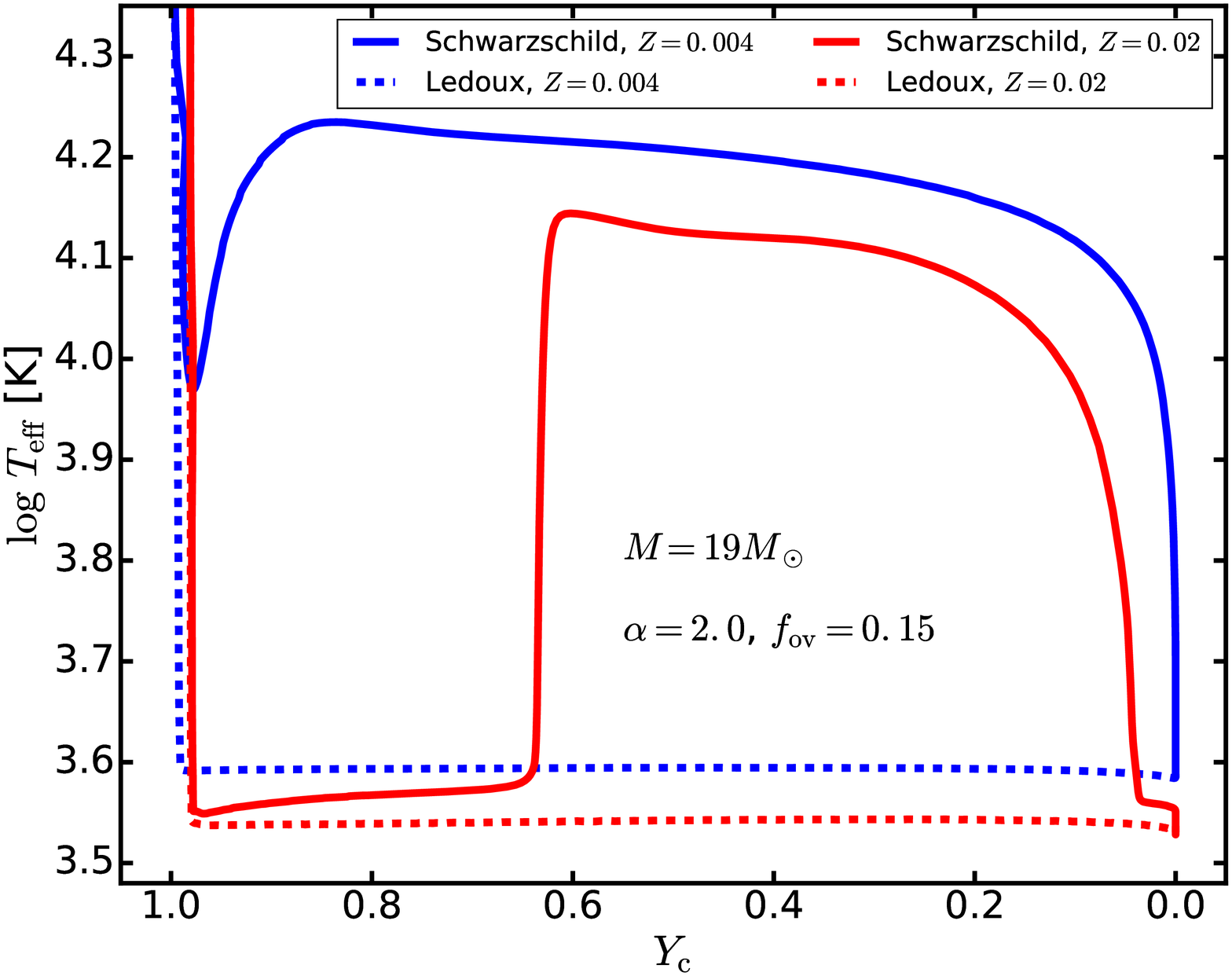}
\caption{Evolution of the effective temperature of a 19~$M_\odot$ star as a function 
of the helium mass fraction at the center during the post-main sequence phase, 
for the metallicities of  $Z = 0.02$ (red) and $Z=0.004$ (blue) and 
the Schwarzschild (solid line) and Ledoux criteria (dashed line) for convection. 
A semiconvection parameter of $\alpha_\mathrm{SEM}$ is used for the Ledoux models. 
\label{fig:Teff_Yc}}
\end{figure}

\begin{table*}
\begin{center}
\caption{The RSG life times and the physical properties at the final state of models with $f_\mathrm{ov}=0.15$}
\label{tab:properties}
\begin{tabular}{ccccccccccccc} 
\hline \hline 
$M_\mathrm{ini}/M_{\sun}$ & \multicolumn{2}{c}{RSG lifetime (yr)} & \multicolumn{2}{c}{$M_\mathrm{f}/M_{\sun}$} & \multicolumn{2}{c}{Log $L_\mathrm{f}/L_{\sun}$} & \multicolumn{2}{c}{Log $T_\mathrm{eff,f}$} 
& \multicolumn{2}{c}{$R_\mathrm{f}/R_{\sun}$} & \multicolumn{2}{c}{$M_\mathrm{H-env}/M_{\sun}$}  \\
\cline{2-3} \cline{4-5} \cline{6-7} \cline{8-9} \cline{10-11} \cline{12-13} 
 & Sch. & Led. & Sch. & Led. & Sch. & Led. & Sch. & Led. & Sch. & Led. & Sch. & Led. \\
\hline
\multicolumn{13}{c}{SMC (Z=0.004), $\alpha=2.0$} \\
\hline 
  11 &  1,650,025 & 1,376,374 & 10.1858 & 10.3023 & 4.7013 & 4.6043 & 3.5875 & 3.5971 &  499.9649  &  427.8429 &  6.8094 &   7.0246  \\
 15 &  990,427   & 746,079 & 12.9335 & 13.1613 & 5.0163 & 4.8982 & 3.5781 & 3.5909 &  750.1762  &  617.4432 &  7.8991 &   8.1447  \\
 19 &  17,514    & 525,039 & 18.3409 & 15.0085 & 5.2281 & 5.1564 & 3.5854 & 3.5814 &  925.7693  &  868.3767 & 11.6041 &   8.0341   \\
 23 &  13,761    & 449,434 & 21.7674 & 15.4192 & 5.3761 & 5.3578 & 3.6153 & 3.5874 &  956.5488  & 1065.0320 & 13.5601 &   6.3494   \\
 27 &  8,200     & 391,866 & 25.2876 & 15.4704 & 5.4887 & 5.5090 & 3.6748 & 3.6516 &  827.9678  &  943.2010 & 14.9002 &   4.2234   \\
 31 &  13,154    & 311,744 & 26.7822 & 16.4931 & 5.5937 & 5.6291 & 3.7125 & 3.7303 &  785.3963  &  753.7852 & 14.6574 &   3.0525   \\
 35 &  2,881     & 202,037 & 30.3458 & 17.6989 & 5.6404 & 5.7304 & 3.7733 & 4.0850 &  626.3249  &  165.4101 & 16.2436 &   2.0399   \\
 39 &  -   & 44,084 & 29.1417 & 20.3055 & 5.8029 & 5.8149 & 3.9018 & 4.1149 &  417.8399  &  158.7948 & 12.9359 &   2.2921   \\
\hline
\multicolumn{13}{c}{LMC (Z=0.007), $\alpha=2.0$} \\
\hline
 11 & 1,671,297   & 1,399,817 & 10.1601 & 10.2269 & 4.6853 & 4.5969 & 3.5741 & 3.5828 &  521.9620 &  452.9277 &   6.8357 &   6.9608     \\
 15 & 994,797     & 743,871 & 12.5099 & 13.0037 & 5.0279 & 4.8940 & 3.5606 & 3.5762 &  824.1839 &  657.4775 &   7.3944 &   7.9884     \\
 19 & 34,388      & 537,440 & 18.0254 & 14.8057 & 5.2171 & 5.1466 & 3.5710 & 3.5662 &  976.9340 &  920.6348 &  11.3898 &   7.8825     \\
 23 & 45,633      & 441,943 & 20.8010 & 15.1176 & 5.3876 & 5.3516 & 3.6069 & 3.5762 & 1007.7337 & 1113.5686 &  12.3493 &   6.0894     \\
 27 & 73,243      & 383,051 & 22.2599 & 15.9264 & 5.5083 & 5.4938 & 3.6669 & 3.6432 &  878.2502 &  963.4212 &  11.8129 &   4.7807     \\
 31 & 6,054       & 313,429 & 25.0192 & 16.2065 & 5.6003 & 5.6247 & 3.7197 & 3.7334 &  765.4984 &  739.3669 &  12.8949 &   2.8254     \\
 35 & 1,523       & 190,926 & 25.7851 & 17.2649 & 5.6462 & 5.7292 & 3.8204 & 4.1127 &  507.7103 &  145.3594 &  11.7467 &   1.6609     \\
 39 & -     & 39,672 & 29.2520 & 19.5958 & 5.7856 & 5.8120 & 3.9049 & 4.1479 &  403.9329 &  135.9745 &  13.3078 &   1.7273     \\
\hline
\multicolumn{13}{c}{Milky Way (Z=0.02), $\alpha=2.5$} \\
\hline
11 &   1,789,498 &      1,523,557 & 10.1744 & 10.1733 & 4.6043 & 4.4795 & 3.5784 & 3.5909 &  466.2985 &  381.2712 &   7.0789 &    7.0877   \\
15 &   1,090,883 &      763,301 & 12.5839 & 12.9326 & 4.9546 & 4.8393 & 3.5667 & 3.5774 &  736.6008 &  613.9191 &   7.8879 &    8.0236    \\
19 &   774,367  &      546,383 & 13.9185 & 14.5915 & 5.1932 & 5.1478 & 3.5598 & 3.5605 & 1000.4361 &  946.4467 &   7.4899 &    7.7396    \\
23 &   603,959  &      448,828 & 16.0096 & 14.9703 & 5.3756 & 5.3306 & 3.6174 & 3.5880 &  946.9883 & 1029.5352 &   7.6622 &    6.0912    \\
27 &   534,168  &      401,404 & 14.7389 & 15.0881 & 5.5115 & 5.4913 & 3.7016 & 3.6618 &  751.1105 &  881.8125 &   4.6147 &    4.0890    \\
31 &   201,536  &      311,624 & 15.2366 & 15.5506 & 5.6370 & 5.6167 & 3.8529 & 3.7473 &  432.4924 &  687.1696 &   3.1200 &    2.4073    \\
35 &   68,392   &      46,487 & 18.8221 & 16.2550 & 5.7035 & 5.7174 & 3.9093 & 4.2900 &  360.1399 &   63.3823 &   4.7451 &    0.8829     \\
39 &   13,397   &      3,139 &     - & 18.0424 &    - & 5.8044 &    - & 4.5100 &       - &   25.4351 &      - &    0.5707      \\
\hline
\multicolumn{13}{c}{M31 (Z=0.04), $\alpha=3.0$} \\
\hline
11 &   1,812,424 &   1,569,048 & 10.1477 & 10.1548 & 4.3479 & 4.3473 & 3.6150 & 3.6119 &  293.2132 &  297.2750  &  7.7263  & 8.5662   \\
15 &   989,166   &   769,008 & 12.5146 & 12.6277 & 4.9716 & 4.8176 & 3.5676 & 3.5856 &  747.8461 &  576.5820  &  7.6327  & 7.7076     \\
19 &   711,926   &   546,657 & 13.1950 & 13.9364 & 5.2232 & 5.1103 & 3.5659 & 3.5687 & 1006.8894 &  873.1242  &  6.2395  & 6.9840     \\
23 &   630,922   &   452,559 & 13.3380 & 13.9021 & 5.3397 & 5.3541 & 3.6171 & 3.6109 &  909.7774 &  951.7694  &  5.2481  & 4.8682     \\
27 &   432,224   &   384,563 & 12.2926 & 14.1260 & 5.5575 & 5.4955 & 3.8079 & 3.7022 &  485.4907 &  735.3421  &  1.1885  & 2.9907     \\
31 &   175,989   &   182,380 &     - & 14.5720 &    - & 5.6159 &    - & 3.8921 &       - &  352.3443  &     -  & 1.3130     \\
35 &   8,241     &   4,002 &     - &     - &    - &    - &    - &    - &       - &       -  &     -  &    -       \\
39 &   527       &   782 &     - &     - &    - &    - &    - &    - &       - &       -  &     -  &    -         \\ 
\hline
\end{tabular}
\end{center}
\end{table*}

The convection criterion becomes particularly relevant in this regard, because
the post main sequence evolution can be significantly affected by its choice.
As shown in Fig.~\ref{fig:Teff_Yc} as an example, with the Ledoux criterion and
slow semiconvection ($\alpha_\mathrm{SEM} = 0.01$), a 19~$M_\odot$ star with
$\alpha = 2.0$ and $f_\mathrm{ov} =0.15$ becomes a RSG right after core
hydrogen exhaustion, and spend the rest of its life as a RSG until the end for
both solar and SMC metallicities. With the Schwarzschild criterion, the stars
of the same parameters undergo a blue loop at solar metallicity, and spend most
of the post-main sequence phase as a blue supergiant (BSG) at SMC metallicity.
RSG lifetimes of some selected models with two different convection criteria
are presented in the Table~\ref{tab:properties}. For SMC and LMC metallicities,
RSG lifetimes become much shorter in the Schwarzschild case than in the Ledoux
case with slow semiconvection. For the metal-rich cases ($Z \ge Z_{\sun}$), RSG
lifetimes with the Schwarzschild criterion are comparable to those of the
Ledoux models.  In general, we find that the BSG to RSG lifetime ratio becomes
higher with a higher mass, lower metallicity and smaller convective
overshooting parameter, when the Schwarzschild criterion is adopted. This is in
qualitative agreement with previous studies on massive stars. In particular,
the number ratio of BSG to RSG stars has been predicted to increase with
decreasing metallicity in the previous stellar evolution models with the
Schwarzschild criterion~\citep[e.g.,][]{Langer1995, Eggenberger2002}.  In
contrast, all our Ledoux models become RSGs shortly after core hydrogen
exhaustion, and none of them undergoes a blue loop phase. As already discussed
in previous studies~\citep{Eggenberger2002}, none of the Schwarzschild and Ledoux
models would be able to explain the observation that the BSG to RSG ratio
increases with increasing metallicity.  The reason for the difference 
between the Schwarzschild and Ledoux cases is difficult to
understand and still a matter of great debate~\citep[e.g.,][]{Alongi1991,
Stothers1991,  El1995, Langer1995, Bono2000}.  In the present study, we focus
our discussion on RSG temperatures and leave the issue of BSG/RSG populations
as a future work. 

Finally, we would like to remind the readers of the fact that MESA adopts the
so-called MLT++ treatment for the energy transport in radiation-dominated
convective regions as a default option for massive stars~\citep{Paxton2013}.
This means that the superadiabaticity is reduced compared to the case of the
standard mixing length theory. This makes the energy transport in the
convective envelopes of RSGs that are close to the Eddington limit significantly more
efficient than in the case of the standard mixing length approximation. RSG
temperatures become higher with MLT++ accordingly when the luminosity is 
sufficiently high ($\log L/L_\odot \gtrsim 5.1$; see the related discussion in
Section~\ref{sec:comp} below).
The physics of the energy transport in radiation-dominated convective regions is poorly understood~\citep[see][for a detailed discussion on this issue]{Paxton2013}.
Note also that the convection in the RSG envelope may become supersonic 
in the outermost layers, which may cause shock energy dissipation, where
the simple approximation of the mixing length theory brakes down. 
Turbulence pressure might also play an important role. 
Therefore, these uncertainties should be
taken into account as a caveat when we compare models with observations.

\section{Comparison with observational data} \label{sec:comp}

In this section, we discuss the metallicity dependence of RSG temperatures
using our MESA models compared to observed RSGs in SMC, LMC, our Galaxy and
M31.  For our discussion below, the RSG effective temperatures inferred from
the strengths of the TiO band \citep[e.g.,][]{Levesque2005, Levesque2006,
Massey2009, Gordon2016, Massey2016} and the spectral energy distribution (SED;
e.g., \citealt{Davies2015, Gazak2015, Patrick2015}) are referred to as TiO and
SED temperatures, respectively.  The bolometric luminosities of the RSGs with
TiO temperatures were adopted from $M_K$ rather than $M_V$ in their
catalogue~\citep{Levesque2005, Levesque2006, Massey2009, Massey2016}, while the
luminosity calibration relations of~\citet{Davies2013} were used for the RSGs
with SED temperatures.

\begin{figure}
\centering
\includegraphics[width=0.9\columnwidth]{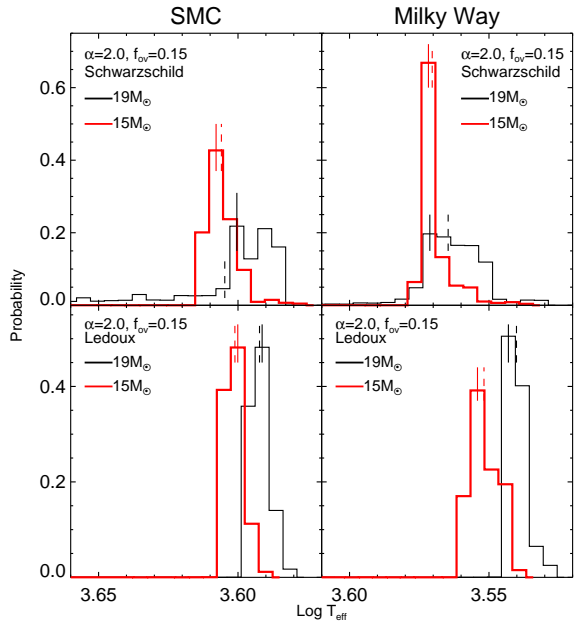}
\caption{
The probability density functions of RSG temperatures for $15$ (red) and $19M_{\sun}$ (black)
models with SMC and Milky Way metallicities in Schwarzschild and Ledoux convection criterion.
The mixing length of $\alpha=2.0$ and overshooting parameter of $f_\mathrm{ov}=0.15$ were adopted. The time-weighted temperatures of $<T_{eff}>$ and 
mode temperatures of $T_{mode}$ are indicated by vertical dashed line and solid line, respectively. 
}
\label{pdf} 
\end{figure}

For the mixing length calibration, we need to determine a representative RSG
temperature at a given  luminosity from our models.  We find that most of the
RSG models have a probability density function (PDF) of RSG temperatures with a well
defined single peak.  This means that the time-averaged RSG temperature
($<T_\mathrm{eff}>_\mathrm{RSG}: = \int_\mathrm{t_\mathrm{RSG}} T_\mathrm{eff}
dt/t_\mathrm{RSG}$) agrees well with the mode value ($T_\mathrm{mode}$) of the
PDF of a given model sequence.  Some exceptions are found for some Schwarzschild
models that deviate from the Hayashi line for a certain fraction of the RSG
phase.  As an example, Figure~\ref{pdf} shows PDF of RSG temperatures for 15~$M_\odot$ and 19~$M_\odot$ models with $\alpha
= 2.0$ and $f_\mathrm{ov} = 0.15$.  The PDFs were derived from the ratios of
$dt$s of temperature bins  to the RSG lifetime ($t_\mathrm{RSG}$). To determine
the RSG lifetime, we adopt $T_\mathrm{eff} < 4800~\mathrm{K}$ ($\log
T_\mathrm{eff} =3.68$) as the criterion for RSGs (i.e., $t_\mathrm{RSG} =
\int_{T_\mathrm{eff} < 4800~\mathrm{K}} dt$), following \citet{Drout2009}.  It
is found that the difference between $<T_\mathrm{eff}>_\mathrm{RSG}$ and
$T_\mathrm{mode}$ is less than about 0.03 dex for $15~M_\odot$ models for both
the Schwarzschild and Ledoux cases.  For 19~$M_\odot$, the difference is as
large as 0.1 dex with the Schwarzschild criterion, while it still remains small
($<$ 0.03 dex) in the Ledoux case.  Such a skewed temperature distribution as in the case of
19~$M_\odot$ Schwarzschild models of the figure is found in particular for
relatively low metallicity (SMC or LMC metallicity) and high initial mass
($M\gtrsim 19~M_\odot$) with the Schwarzschild criterion.  In this case, the RSG
temperature of a given model sequence would be better represented by
$<T_\mathrm{eff}>$ than $T_\mathrm{mode}$. We therefore decide to use the
time-averaged temperature ($<T_\mathrm{eff}>_\mathrm{RSG}$)  and luminosity
($<L_\mathrm{eff}>_\mathrm{RSG}$) for our mixing length calibration.

More specifically, using our model results, we interpolate
$<T_\mathrm{eff}>_\mathrm{RSG}$ and $<L>_\mathrm{RSG}$ at mixing length values
from $\alpha =$ 1.5 to 3.0 with $0.1$ increment, for a given set of the
convection criterion, $f_\mathrm{ov}$, and metallicity.  Then, we compare the
temperatures of observed RSGs with those of interpolated values for a given
luminosity. The deviation between the observations and the model temperatures
is used to compute a $\chi^2$ value.  The mixing length value that gives the
lowest $\chi^2$ is determined to be our calibrated value that can best
reproduce the observed RSG temperatures. 

As discussed above, the effect of convective overshooting on RSG temperatures
is minor compared to that of the mixing length. Following \citet{Martins2013},
we take $f_\mathrm{ov} = 0.15$ as the fiducial value in our discussion below.

\begin{figure*}
\centering
\includegraphics[width=0.9\textwidth]{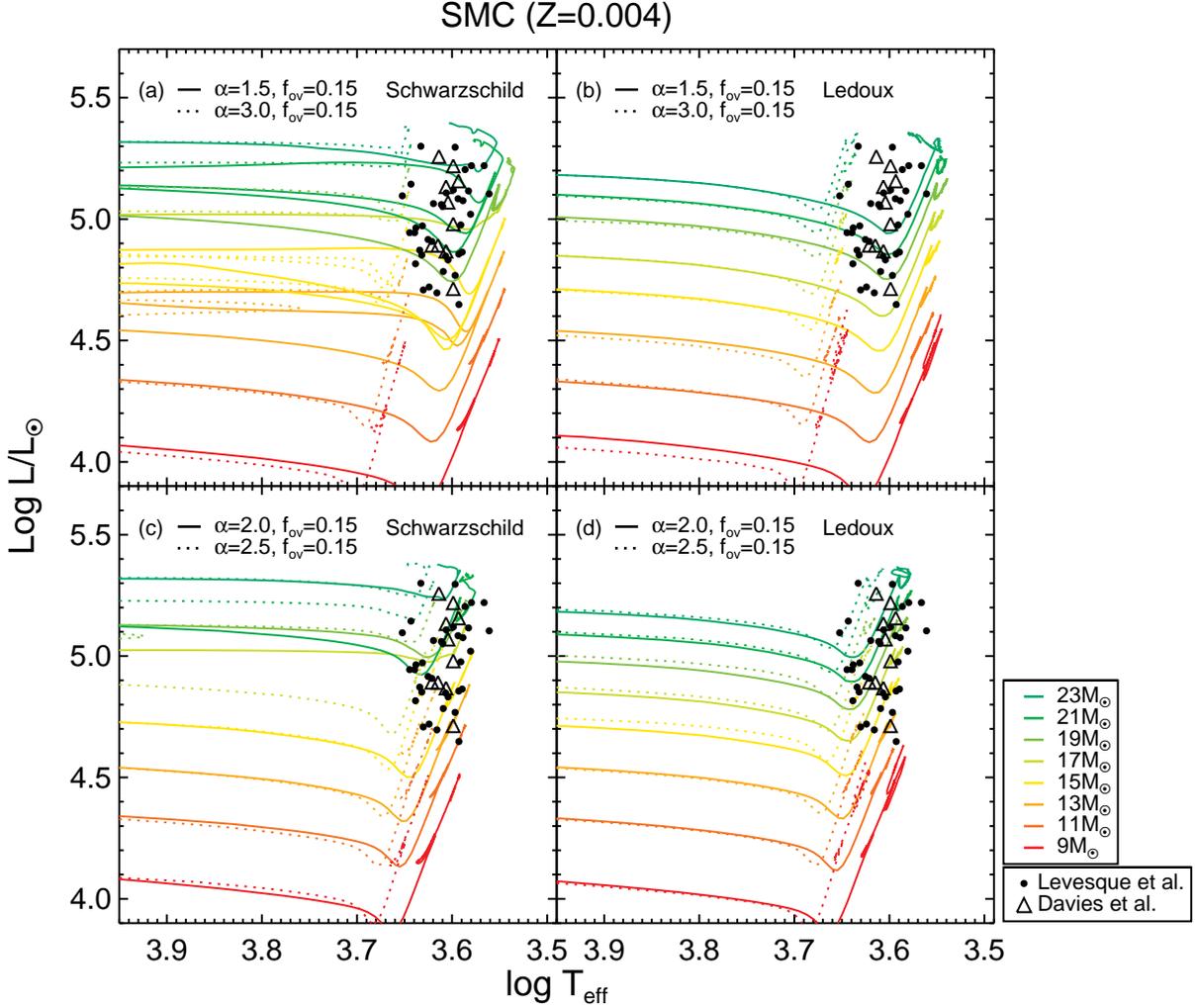}
\caption{
Evolutionary tracks on the HR diagram of the SMC-like metallicity ($Z=0.004$)
models with $f_\mathrm{ov} = 0.15$.  The Schwarzschild and Ledoux models are
presented in the left and right panels, respectively. The tracks with $\alpha =
1.5$ (solid line) and $\alpha = 3.0$ (dotted line) are given in the upper
panels, and those with $\alpha = 2.0$ (solid line) and $\alpha = 2.5$ (dotted line) in the
lower panels.  The initial mass of each track is indicated by the color of the line.  The
RSG samples of the SMC from~\citet[TiO temperatures]{Levesque2006} and from~\citet[SED temperatures]{Davies2015}
are indicated by black dots and open triangles, respectively.  
}\label{smc} 
\end{figure*} 

\begin{figure*}
\centering
\includegraphics[width=0.9\textwidth]{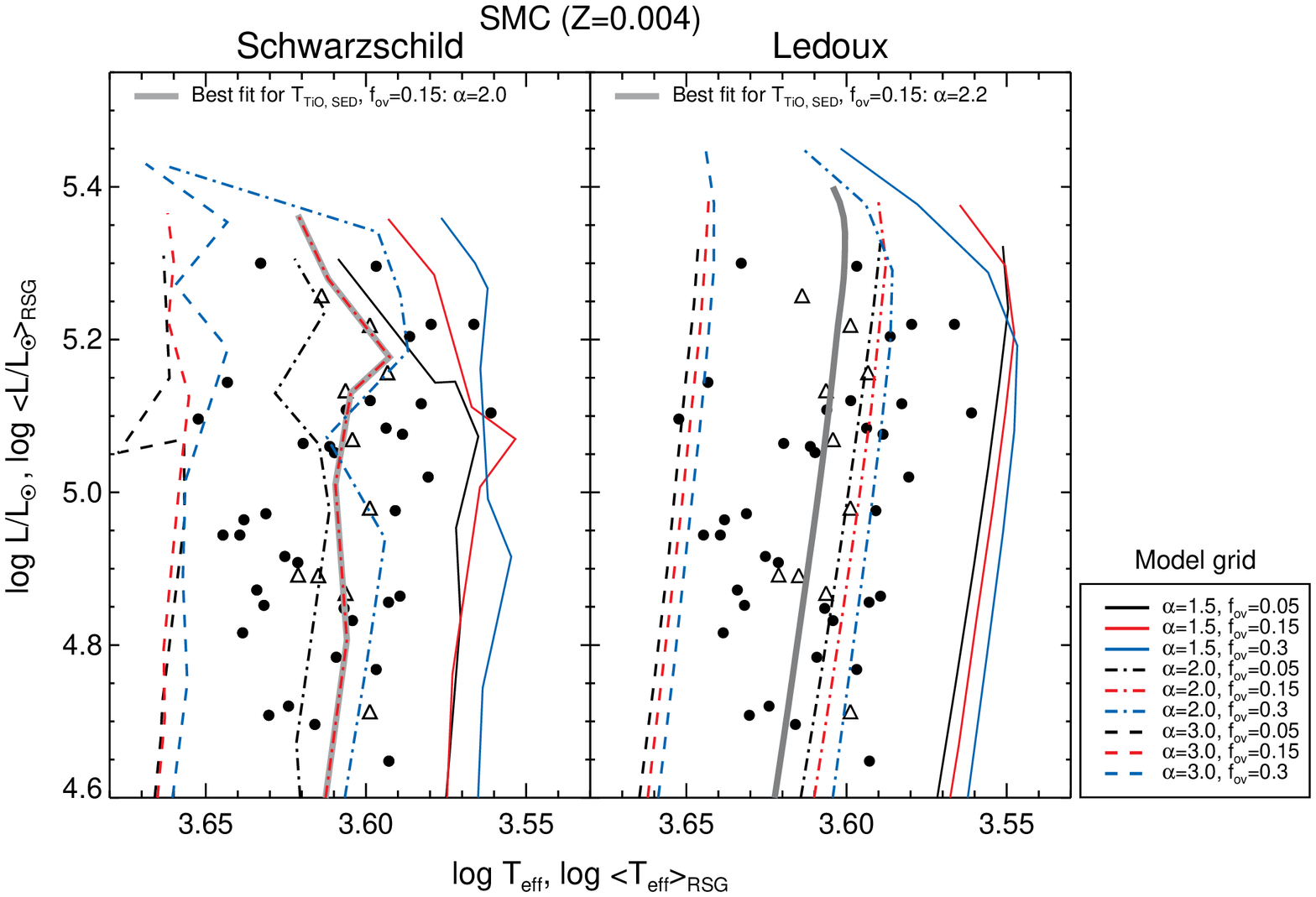}
\caption{The time-weighted temperatures and luminosities (see Eqs.~(1) and (2)) of the SMC-like
metallicity ($Z=0.004$) evolutionary tracks on the HR diagram, compared to the observed RSG samples 
of~\citet[TiO temperatures; filled circles]{Levesque2006} and~\citet[SED temperatures; open triangles]{Davies2015}. 
The derived
time-weighted values for three overshooting parameters ($f_\mathrm{ov}=0.05, 0.15$ and
$0.3$) are represented by black, red, and blue lines, respectively.  The
results for three mixing length parameters of $\alpha=1.5, 2.0$ and $3.0$ are
indicated by solid, dot-dashed, and dashed lines, respectively.  
The best fits to TiO and SED temperatures, which are obtained with 
the models using  $f_\mathrm{ov} = 0.15$, are indicated by the thick grey lines. 
The corresponding 
calibrated mixing length values are $\alpha = 2.0$ and $\alpha = 2.2$
for the Schwarzschild and Ledoux cases, respectively. 
In the SMC, 
the best fit line is the same for both TiO and SED temperatures.}\label{SMCexp} 
\end{figure*}

\subsection{Small Magellanic Cloud (Z=0.004)}

We present the MESA evolutionary tracks with $f_\mathrm{ov} = 0.15$  at
SMC-like metallicity (Z=0.004) on the HR diagram compared with the observed SMC
RSGs of \citet{Levesque2006} and \citet{Davies2015} in Figure~\ref{smc}.
Although SED temperatures are known to be systematically higher than TiO
temperatures~\citep{Davies2013}, such an offset is not found with the SMC
samples of the figure.  The TiO temperatures given by~\citet{Levesque2006} are
more widely spread on the HR diagram than the SED temperatures of
\citet{Davies2015}. See Section~\ref{sec:calibration} below for a related
discussion. 

We find that RSG models with the mixing length of $\alpha=2.0$ and $\alpha =
2.5$ can roughly reproduce both the TiO and SED temperatures.  RSG models with
$\alpha=1.5$ and $3.0$ are too cool and too warm, respectively, compared to the
observed RSGs.  On the other hand, \citet{Patrick2015} found that RSG models of
the Geneva group~\citep{Georgy2013} where $\alpha = 1.6$ are adopted are
systematically warmer than the RSGs  of~\citet{Davies2015}. This discrepancy
between our models and Geneva models is because the Geneva code gives
systematically higher RSG temperatures for a given mixing length than the MESA
code does (see the discussion in Section~\ref{sec:effects}) and because the
models by \citet{Georgy2013} have a lower metallicity ($Z=0.002$) than the
value adopted in our models (i.e., $Z=0.004$) that is typically invoked for the
SMC.

In Figure~\ref{SMCexp}, we present the time-weighted temperatures and
luminosities of our RSG models on the HR diagram as well as best-fitted values
to the observed RSG temperatures in the SMC. The observed RSGs are within the
boundaries provided by the models with $\alpha = 1.5$ and $\alpha = 3.0$.  In
both the Schwarzschild and Ledoux cases, the effect of overshooting on RSG
temperatures is minor compared to the effect of the mixing length.  For a given
mixing length, the time-weighted temperatures are slightly lower for the Ledoux
models than for the Schwarzschild models.  This is partly because of the fact
that many of the Schwarzschild models at the SMC metallicity tend to deviate
from the Hayashi line for a significant fraction of the RSG phase, while the
Ledoux models remain on the Hayashi line for almost all of the RSG phase as
explained in Sect.~\ref{sec:effects}. 

We find that with the Schwarzschild models and $f_\mathrm{ov}=0.15$,  $\alpha = 2.0$
gives the best fits to the data for both TiO and SED temperatures.  The
corresponding values with the Ledoux models are $\alpha = 2.2$.  These values
become somewhat lower/higher for a smaller/larger overshooting parameter, as
discussed in Sect.~\ref{sec:calibration} below. 
We conclude that $\alpha \simeq 2.0$ can result in RSG models that
can provide a reasonably good fit to RSG temperatures in the SMC.

\begin{figure*}
\centering
\includegraphics[width=0.9\textwidth]{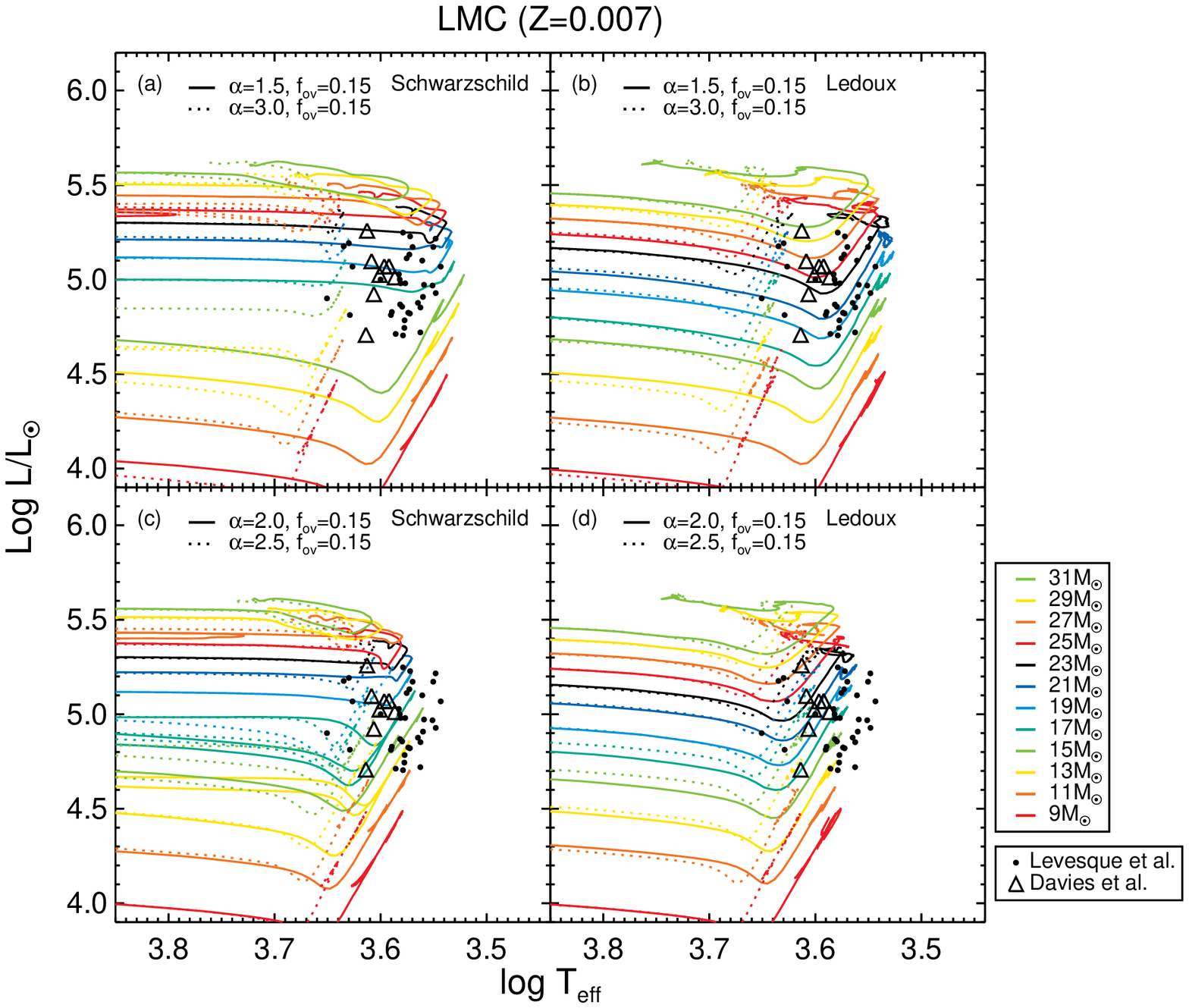}
\caption{Same as in Figure~\ref{smc}, but for the LMC-like metallicity ($Z = 0.007$). }
\label{lmc}
\end{figure*}

\begin{figure*}
\centering
\includegraphics[width=0.9\textwidth]{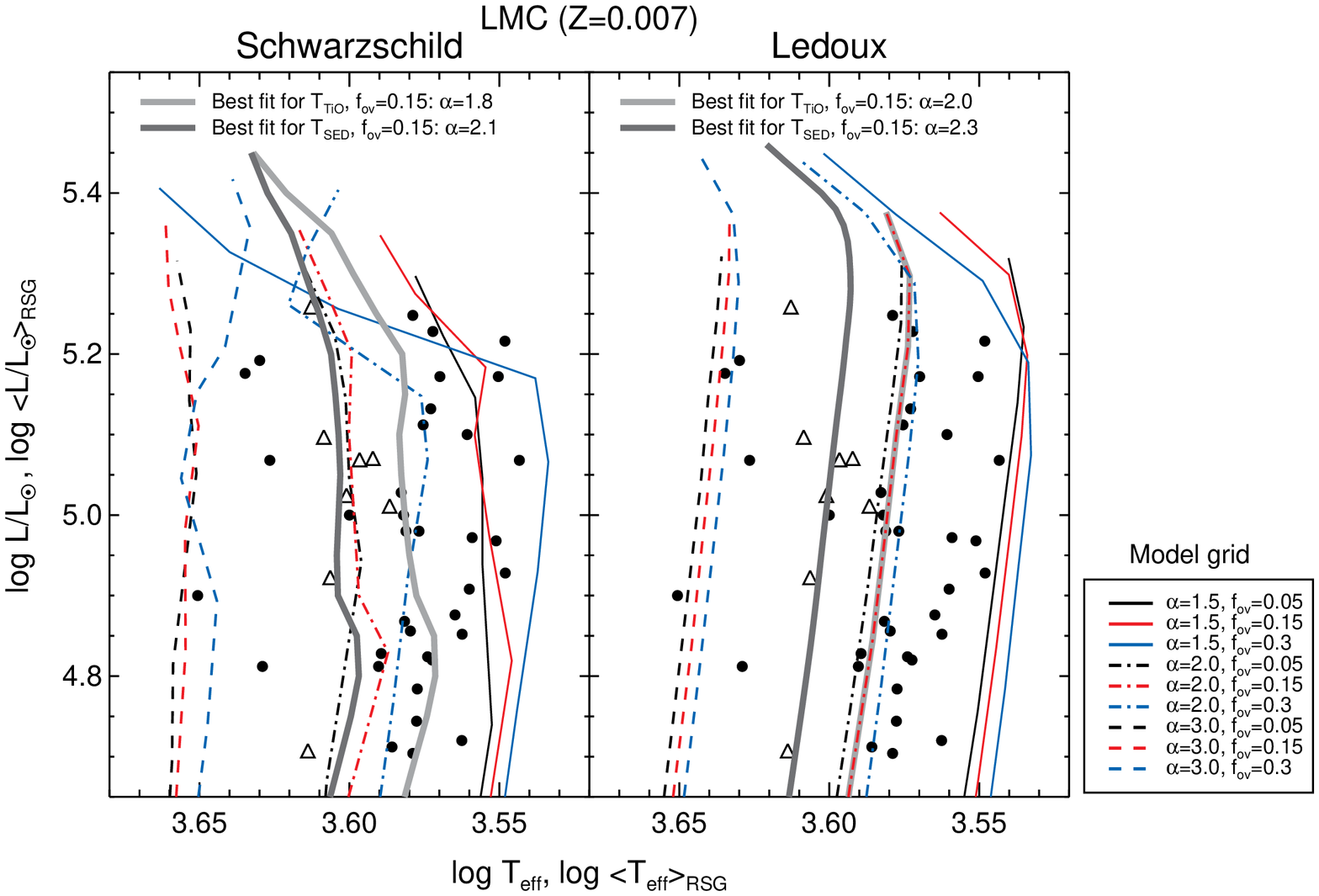}
\caption{
Same as in Fig.~\ref{SMCexp} but for the LMC-like metallicity ($Z=0.007$). 
The best fit lines obtained from the models with $f_\mathrm{ov}=0.15$ for the TiO and SED temperatures  
are marked by the light and dark grey lines, respectively.
}
\label{LMCexp}
\end{figure*}

\subsection{Large Magellanic Cloud (Z=0.007)}

Figure~\ref{lmc} shows the MESA evolutionary tracks at the LMC-like metallicity
(Z=0.007) and observed RSGs in the LMC from~\citet{Levesque2006} and
~\citet{Davies2015}.  In contrast to the SMC
case, the SED temperatures of~\citet{Davies2013} are systematically higher than
the TiO temperatures of~\citet{Levesque2006} in the LMC, although there exists a
significant overlap.  We find that the RSGs models with $\alpha=1.5$ and $3.0$
have significantly lower and higher RSG temperatures compared to the
observations, respectively.  Models with $\alpha=$ 2.0 and 2.5 (dotted lines)
can roughly reproduce both the TiO and SED temperatures, as in the case of the SMC. 

In Figure~\ref{LMCexp}, we present the time-weighted temperatures and
luminosities of our model grids at LMC-like metallicity on the HR diagram,
compared with the observed RSGs.  We find that the best fit values of $\alpha$
for the TiO temperatures are smaller than those for the SED temperatures: with
$f_\mathrm{ov}=0.15$, $\alpha =$ 1.8 (TiO) and 2.1 (SED) for the Schwarzschild
models and $\alpha =$ 2.0 (TiO) and 2.3 (SED) for the Ledoux models,
respectively. This confirms the systematic offset between TiO and SED
temperatures discussed by \citet{Davies2013}.

\begin{figure*}
\centering
\includegraphics[width=0.9\textwidth]{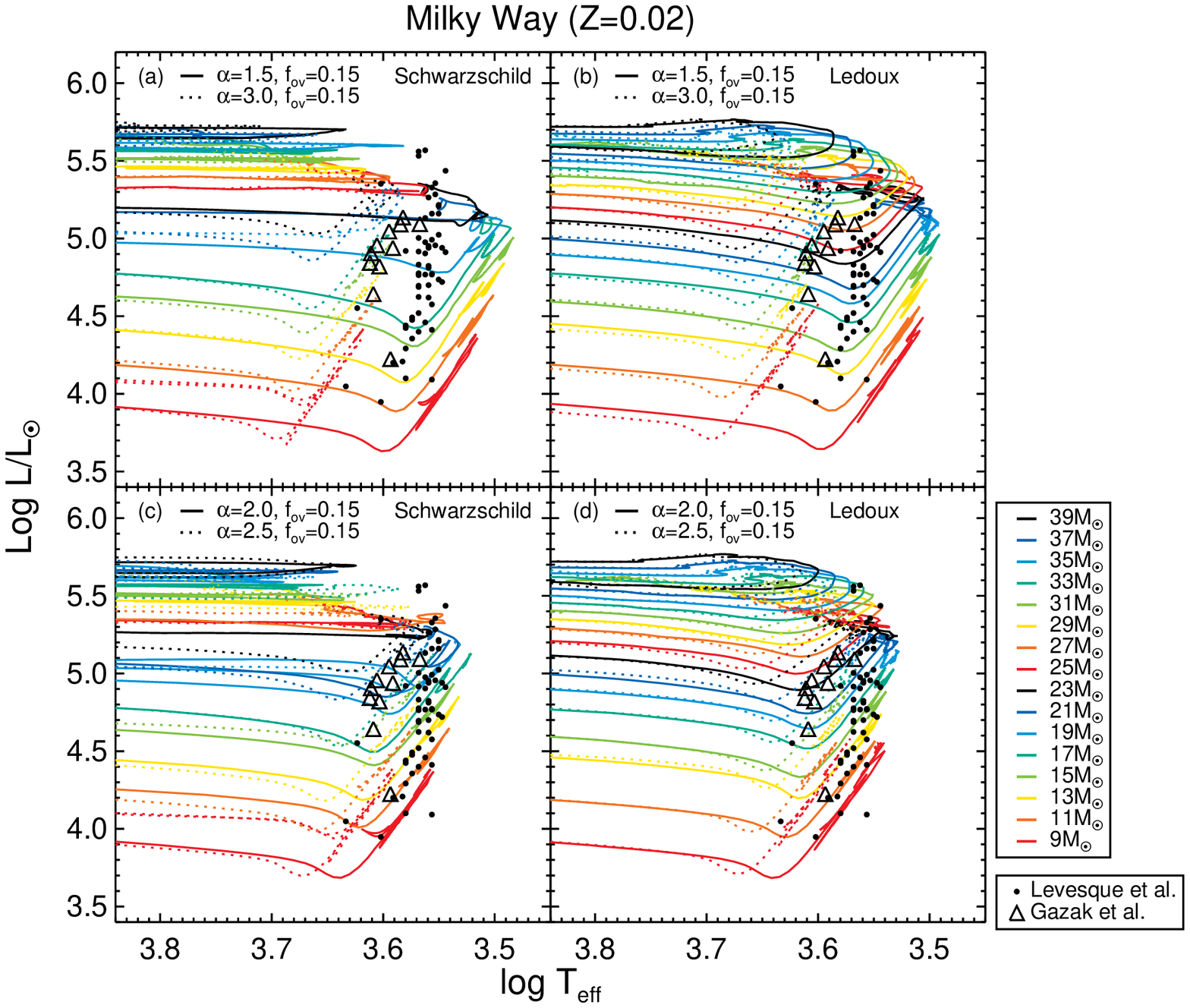}
\caption{
Same with Figure~\ref{smc}, but for solar metallicity ($Z=0.02$). 
The compared Galactic RSG samples are taken from  \citet[TiO temperatures; filled circles]{Levesque2005} and \citet[SED temperatures; open triangles]{Gazak2014}. 
}
\label{milky}
\end{figure*}

\begin{figure*}
\centering
\includegraphics[width=0.9\textwidth]{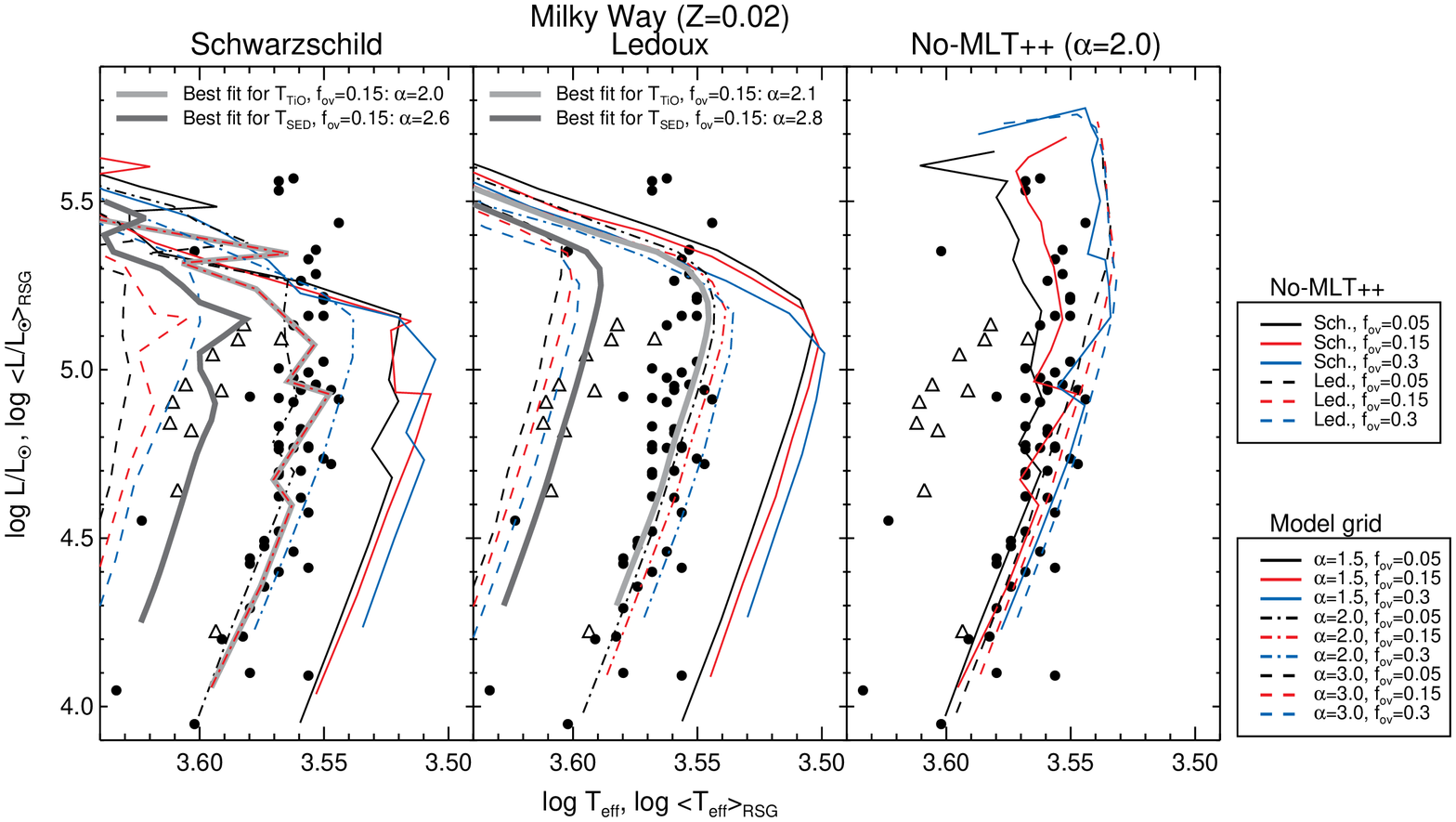}
\caption{Same with Figure~\ref{SMCexp}, but for solar metallicity ($Z=0.02$). 
The results without the MLT++ treatment and with
$\alpha=2.0$ are also plotted in third panel for comparison.  The compared
Galactic RSG samples are taken from ~\citet[TiO temperatures; filled
circles]{Levesque2005} and~\citet[SED temperatures; open triangles]{Gazak2014}. 
}
\label{MKexpand}
\end{figure*}

\subsection{Milky Way (Z=0.02)}

In Figure~\ref{milky}, we compare the MESA evolutionary tracks with
$f_\mathrm{ov} = 0.15$ at solar metallicity (Z=0.02) and the observed Galactic
RSGs of~\citet{Levesque2005} and~\citet{Gazak2014} on the HR diagram.  As shown
in Figure~\ref{milky}, the SED temperatures by~\citet{Gazak2014} are
significantly higher than TiO temperatures by~\citet{Levesque2005} for the RSGs
in the Milky Way. We find that the evolutionary tracks with $\alpha=2.0$ and
$\alpha=2.5$ are roughly compatible with the positions of the observed RSGs
from~\citet{Levesque2005} and ~\citet{Gazak2014}, respectively.  The
temperatures of the RSG models with $\alpha=1.5$ are too low to reproduce the
observed RSGs.  

The time-weighted temperatures and luminosities of the RSGs models at
solar-metallicity are shown in Figure~\ref{MKexpand}.  The lines with
$\alpha=2.0$ agree well with the TiO temperatures for both the Schwarzschild
and Ledoux cases.  The SED temperatures of~\citet{Gazak2014} are in-between the
lines of  $\alpha=2.0$  and $3.0$.  For $f_\mathrm{ov} = 0.15$, the best fits of
the Schwarzschild models to the observations are found with $\alpha=2.0$ and
$2.6$ for TiO and SED temperatures, respectively.  With the Ledoux models,
$\alpha=2.1$ and $2.8$ give the best fits to the TiO and SED temperatures. 

Note that the most luminous Galactic RSGs with $ \log L/L_\odot \gtrsim 5.3$
are much cooler compared to our model predictions for all our considered mixing
lengths and overshooting parameters. RSGs models tend to have lower
temperatures for higher luminosities for $ \log L/L_\odot \lesssim 5.3$, which
roughly agrees with observations. However, the $<T_\mathrm{eff}> - <L>$ lines
begin to bend towards the left as the luminosity increases beyond $\log
L/L_\odot \approx 5.3$. This is caused mainly by the MLT++ treatment
of MESA that leads to more efficient energy transport compared to the case of
the ordinary mixing-length formulation (see the discussion in
Section~\ref{sec:effects}).  
This makes very luminous RSGs quickly move away from the Hayashi line.
For comparison, we present the model results for
which the MLT++ option is turned off in Figure~\ref{MKexpand} (the third
panel). The bending toward the left for $\log L/L_\odot \gtrsim 5.3$ is still
found because of strong mass loss from such luminous RSGs, but its degree is
much weaker than in the case of MLT++. The temperatures of the most luminous
RSGs are better matched by the models without MTL++, implying  that the MLT++
treatment requires a caution when applied to the most luminous RSGs.  To
understand this discrepancy, we should address both the validity of the MLT++
treatment and the uncertainty in the temperature estimates of these most luminous
RSGs that suffer strong reddening due to circumstellar
dusts~\cite[e.g.,][]{Levesque2005, Gordon2016}. 
The number of these luminous RSGs is small, and this bending with MLT++ does not affect
our result of mixing length calibration.

\subsection{M31 (Z=0.04)}
\begin{figure*}
\centering
\includegraphics[width=0.9\textwidth]{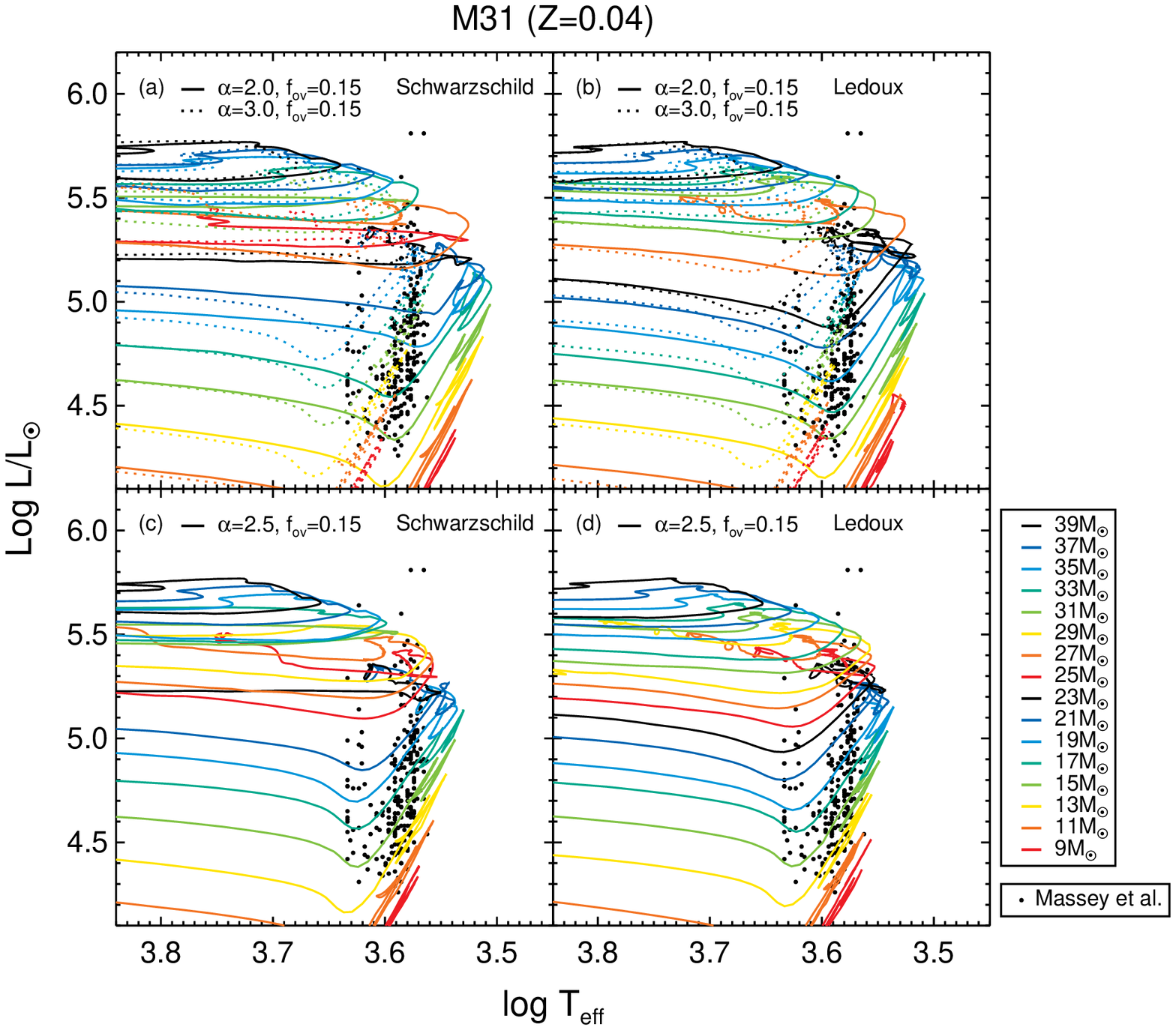}
\caption{The M31-like metallicity (Z=0.04) evolutionary tracks on the HR diagram compared with 
the observed M31 RSG sample of~\citet{Massey2016}. The tracks with $\alpha =
2.0$ (solid line) and $\alpha = 3.0$ (dotted line) are given in the upper
panels, and those with $\alpha = 2.5$ (solid line) in the lower panels.
}
\label{M31}
\end{figure*}

\begin{figure*}
\centering
\includegraphics[width=0.9\textwidth]{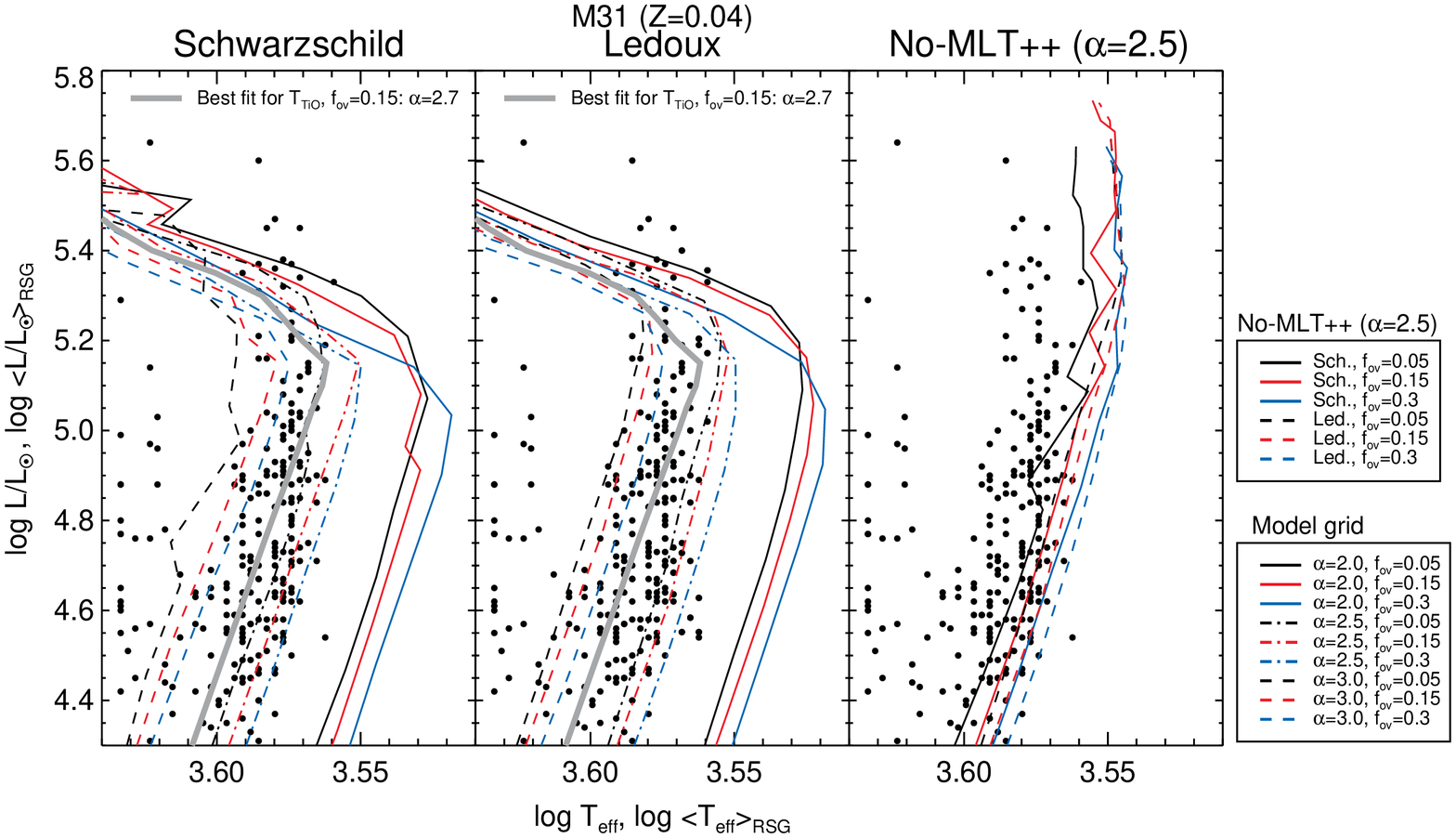}
\caption{The time-weighted temperatures and luminosities of the M31-like
metallicity (Z=0.04) evolutionary tracks of the Schwarzschild (the first panel)
and Ledoux (the second panel) models.  The M31 RSG sample
of~\citet[TiO temperatures]{Massey2016} are marked by filled circles.  The
results without the MLT++ treatment and with $\alpha=2.5$
are presented in the third panel for comparison.  The models of three
overshooting parameters ($f_\mathrm{ov}=0.05, 0.15$ and $0.3$) are represented by black,
red, and blue lines, respectively.  The results of three mixing length
parameters of $\alpha=2.0, 2.5$ and $3.0$ are plotted by solid, dot-dashed, and
dashed lines, respectively.  The light grey lines in the first and second panels
are the best fit lines obtained from the models with $f_\mathrm{ov} = 0.15$. 
\label{M31exp}}
\end{figure*}

In Figure~\ref{M31}, we show the M31-like metallicity $(Z=0.04)$ evolutionary
tracks with $f_\mathrm{ov} = 0.15$ on the HR diagram, compared with the RSG
sample of M31 provided by~\citet{Massey2016} who obtained the RSG temperatures
using the TiO band.  SED temperatures of RSGs in M31 are not available yet.  

The most striking feature of the observed RSGs in M31 is the bifurcation in the
temperature distribution: the warm sequence  at around $\log T_\mathrm{eff} =
3.63$ and  the cool sequence at $\log T_\mathrm{eff} =$ 3.57 --  3.61.
\citet{Massey2016}  investigated the lifetimes of RSG models at solar
metallicity given by \citet{Ekstrom2012} as a function of the effective
temperature.  They found a lifetime gap for the temperature range of
$4100-4150$ K only for the $M = 25~M_\odot$ model sequence, and the bifurcation
is not predicted by lower mass models.  We could not find a lifetime gap at
this temperature range with our models either.  All of our RSG models at the
M31 metallicity stay in the temperature range of $3900-4300$ K only for a very
short time (i.e., less than thousands of years), and spend the most of the RSG
phase at lower temperatures.  

As shown in the figure, the RSG models with $\alpha=2.0$ are too cool to
explain the observations. The tracks with $\alpha = 2.5$ can roughly
reproduce the location of  the RSGs of the cool sequence.  The RSGs of the warm
sequence are too hot to be matched with our RSG models, even with the largest
mixing length value (i.e., $\alpha = 3.0$).  

In Figure~\ref{M31exp}, we present the time-weighted temperatures and
luminosities of our evolutionary tracks at M31-metallicity, compared with
observations. It is clearly seen that the RSG temperatures of the warm sequence
cannot be explained with our considered range of mixing length values.  We
would need a significantly larger value of $\alpha$ than 3.0, or a lower
metallicity to match the temperatures of the warm sequence.  Given that the
physical origin of this warm sequence is not clear  and that our
models do not predict the warm sequence, we calibrate $\alpha$ only with the
RSGs of the cool sequence.  We find that $\alpha = 2.7$ gives the best fits for
both the Schwarzschild and Ledoux models. This is significantly larger than
those found with the TiO data of the other galaxies.  

As in the case of the Milky Way, the inclusion of MLT++ tends to make very
luminous RSG models ($\log L/L_\odot \gtrsim 5.2$) warmer than  those without
MLT++ (compare the first and third panels of Figure~\ref{M31exp}) but
does not affect our mixing length calibration because of the small number
of observed RSGs with $\log L/L_\odot > 5.2$.

\subsection{Discussion}\label{sec:calibration}
\begin{figure}
\centering
\includegraphics[width=0.9\columnwidth]{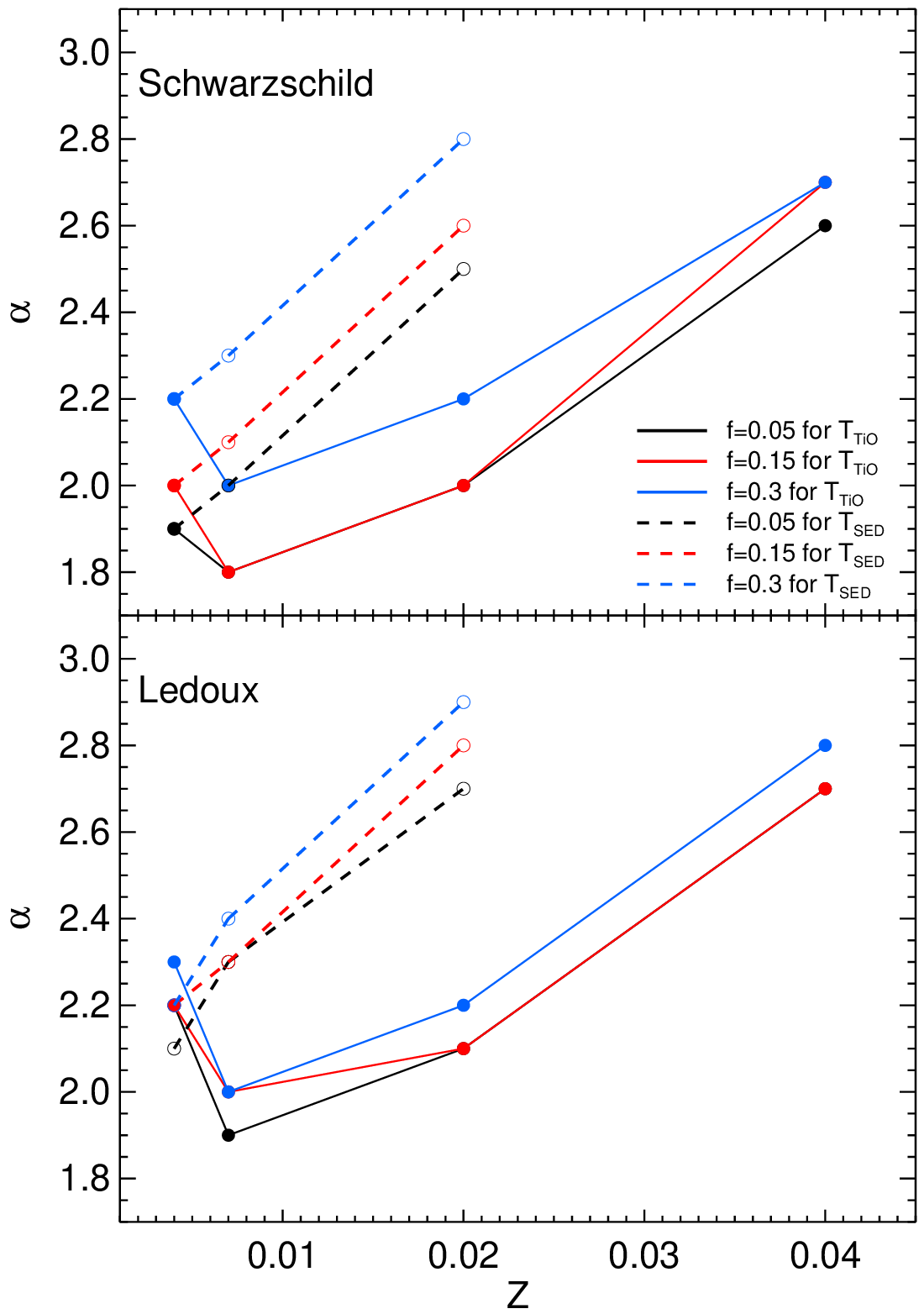}
\caption{The calibrated mixing length values for TiO (solid lines) and SED (dashed lines) temperatures 
as a function of metallicity obtained
with the Schwarzschild (upper panel) and
Ledoux (lower panel) models. 
The adopted overshooting parameters in the models are indicated by three different colors: black ($f_\mathrm{ov}=0.05$), red ($f_\mathrm{ov}=0.15$), 
and blue ($f_\mathrm{ov}=0.30$).} 
\label{mixing}
\end{figure}

We have compared our evolutionary models with observed RSGs of several
different metallicities and calibrated the mixing length for each metallicity.
The result is summarized in Table~\ref{tab1} and Figure~\ref{mixing} where  we
present the calibrated mixing length values for three different overshooting
parameters ($f_\mathrm{ov} =$ 0.05, 0.15 and 0.30), for both the TiO and SED
temperatures.  As shown above, the time-weighted temperatures are
systematically lower for a larger $f_\mathrm{ov}$ and the resultant calibrated
mixing length values are systematically larger for a larger $f_\mathrm{ov}$.
The mixing length values from SED temperatures are higher than those from TiO
temperatures for LMC and Milky Way metallicities, as expected from the fact
that SED temperatures are systematically higher than TiO temperatures.  At
SMC-like metallicity, the difference between the two cases is minor. 

From both TiO and SED temperatures we find strong evidence that the mixing
length depends on metallicity. With TiO temperatures, the metallicity
dependence is particularly evident with the M31 sample of \citet{Massey2016}.
Interestingly, the TiO temperatures of the M31 RSGs of the cool sequence
appears to be systematically higher than those of Galactic RSGs.  Even if we
use the $Z=0.02$  models instead of $Z=0.04$ models for the mixing length
calibration of the M31 sample, we get $\alpha =$ 2.3 and 2.4 for the
Schwarzschild and Ledoux cases with $f_\mathrm{ov} = 0.15$, respectively, which
are significantly larger than  the values of $\alpha =$ 2.0 and 2.1 obtained
with the TiO temperatures of the Galactic RSGs (Table~\ref{tab1}).  This cannot
be easily explained without invoking a metallicity-dependent mixing length,
given that the average metallicity of M31 RSGs is likely to be significantly
higher than the Galactic value ~\citep[see, however,][]{Sanders2012}. 

The  mixing length calibrated with TiO temperatures continuously decreases as
the metallicity decreases from $Z=0.04$ (M31) to $Z=0.007$ (LMC), and suddenly
increases at $Z=0.004$ (SMC). This anomalous behavior at SMC-like metallicity
is related to the wide spread of TiO temperatures of RSGs in
SMC~(Figure~\ref{SMCexp}).~\citet{Levesque2006} argued that this large spread
results from enhanced effects of rotationally induced chemical mixing at
relatively low metallicity of SMC. However, the temperature discrepancy during
the RSG phase between non-rotating and rotating cases is not clearly seen in
recently published stellar evolution models~\citep{Brott2011, Georgy2013}.
This scenario needs to be tested with a large grid of RSG models for a wide
range of initial rotation velocities, which is beyond the scope of the present study. 

The SED temperature ranges are much narrower than those of TiO temperatures
(Figures~\ref{SMCexp}, \ref{LMCexp}, and \ref{MKexpand}), given the small size of
the selected RSG sample of \citet{Davies2015}. In addition, the SED temperature
range does not appear to depend on
metallicity~\citep{Davies2015,Gazak2015,Patrick2015}.  As a result, the
calibrated mixing length with SED temperatures is found to be a monotonically
decreasing function of metallicity and its metallicity dependence appears to be
stronger than in the case with TiO temperatures. 

Interestingly,~\citet{Tayar2017} has also found evidence for
a metallicity-dependent mixing length in Galactic red giant stars, by analyzing
the APOGEE-\emph{Kepler} data. They calibrated the mixing length using 
low-mass star models ($0.6 M_\odot$ -- $2.6 M_\odot$) for a metallicity range
of $[Fe/H] = -2.0 \sim +0.6$, and concluded that the mixing length should be
systematically smaller for lower metallicity (i.e., $\delta \alpha \approx$ 0.2
per dex in metallicity)  to match the temperatures of the observed red giant
stars. Some evidence of the metallicity-dependent mixing length for red giants was also reported 
by~\citet{Chieffi1995}. These qualitatively conform to our finding with RSGs, and  seems to
indicate that less efficient convective energy transport at lower metallicity
is a universal property of the convective envelopes of post-main sequence stars
for both low and high masses. 

This contradicts the theoretical result of \citet{Magic2015} who found that the
mixing length increases with decreasing metallicity in three-dimensional
numerical simulations\footnote{\citet{Stothers1996} also previously suggested
similar conclusions of~\citet{Magic2015} 
based on the old observational data}.
Note, however, that these simulations
focused on stars with higher temperatures and gravities than those of RSGs and
cannot be directly compared to our result. 
The  mixing length theory has limitations to describe 
the RSG convection which can be supersonic in the outermost layers of the envelope.
To our knowledge, there have been no
theoretical studies using multi-dimensional numerical simulations done yet on
the metallicity dependence of the convective energy transport in RSGs, and this
should be an important subject of future studies. 

It is also noteworthy that the discrepancy between the TiO and SED calibration
values (i.e., $\Delta \alpha = \alpha_\mathrm{SED} - \alpha_\mathrm{TiO}$) is
larger for higher metallicity. For example, with $f_\mathrm{ov} = 0.15$  and
the Ledoux criterion, we have $\Delta \alpha =$ 0.0, 0.3, and 0.7 for SMC, LMC
and Milky Way metallicities, respectively.  This might imply that the layer
suitable for the formation of the TiO band  is located systematically  farther
above the continuum photosphere for higher
metallicity~\citep[cf.][]{Chiavassa2011, Davies2013}. However, the size of the
SED samples is much smaller than that of the TiO samples, and the selection
bias might be an alternative reason for this tendency of increasing $\Delta
\alpha$ with metallicity. 

\begin{table*}
\begin{center}
\caption{Calibrated mixing length $\alpha$}\label{tab1}
\begin{tabular}{l  c c c  c c c  c c c  c c c} 
\hline \hline
 & \multicolumn{6}{c}{TiO} & \multicolumn{6}{c}{SED} \\
 & \multicolumn{3}{c}{Schwarzschild} & \multicolumn{3}{c}{Ledoux} & \multicolumn{3}{c}{Schwarzschild} & \multicolumn{3}{c}{Ledoux} \\ 
 $~~~~~~~~f_\mathrm{ov} = $ & 0.05 & 0.15 & 0.30 & 0.05 & 0.15 & 0.30 & 0.05 & 0.15 & 0.30 & 0.05 & 0.15 & 0.30  \\
\hline
SMC ($Z=0.004$) &  1.9  & 2.0 &  2.2  & 2.2 & 2.2 & 2.3 & 1.9 & 2.0 & 2.2 & 2.1 & 2.2 & 2.2 \\ 
LMC ($Z=0.007$) &  1.8  & 1.8 &  2.0  & 1.9 & 2.0 & 2.0 & 2.0 & 2.1 & 2.3 & 2.3 & 2.3 & 2.4 \\
MW ($Z=0.02$)   & 2.0 &  2.0  & 2.2 & 2.1 & 2.1 & 2.2 & 2.5 & 2.6 & 2.8 & 2.7 & 2.8 & 2.9 \\ 
M31    ($Z=0.04$)      & 2.6 & 2.7 & 2.7 & 2.7 & 2.7 & 2.8 &  -    &  -  &  -  &  -  & -   & - \\
\hline
M31    ($Z=0.02$)$^{(a)}$      & 2.2 & 2.3 & 2.4 & 2.4 & 2.4 & 2.5 &  -    &  -  &  -  &  -  & -   & - \\
\hline
\end{tabular}
\tablecomments{(a) Solar metallicity models are used for the mixing length calibration with the M31 RSG sample.} 
\end{center}
\end{table*}

\section{Implications for Type IIP supernova progenitors}\label{sec:final}

Here we discuss the implications of our mixing-length calibration result for SN
IIP supernova progenitors.  
For this purpose, in the figures of Appendix A
(Figures~\ref{SMCrad}, \ref{LMCrad}, \ref{MKrad}, and \ref{M31rad}), we present
the  final radius, total  mass $M_\mathrm{final}$,  and hydrogen envelope mass
$M_\mathrm{H-env}$ at the final evolutionary stage, which we obtain by
interpolating the results of our last computed models for our calibrated
mixing length parameters.  
Examples of physical structures of our last computed models are indicated in Table~\ref{tab:properties}.

\subsection{Final and hydrogen envelope masses}\label{sec:masses}
As shown in Figure~\ref{fig:code}, different choices of the mixing length
within our considered parameter space can hardly alter the evolution on the
main sequence.  However, the role of the mixing length on the mass-loss history
during the post main sequence evolution is significant because the mass-loss
rate depends on the effective temperature of a star, as well as its luminosity.
In our models, the mass-loss rate prescription for RSGs by \citet{deJager1988}
is adopted, which has the power-law relation of $\dot{M} \propto L^{1.769}
T_\mathrm{eff}^{-1.676}$. Given that RSG models with $\alpha = $ 1.5 and  3.0
have a temperature difference of about 0.1 dex on average
(Figures.~\ref{SMCexp}, \ref{LMCexp}, and \ref{MKexpand}), a smaller mixing
length parameter leads to more mass loss. For example, our Ledoux models at
solar metallicity with $M_\mathrm{init}=25~ M_\odot$ and $f_\mathrm{ov} = 0.15$
have final masses of 14.1~$M_\odot$ and 15.6~$M_\odot$ for $\alpha =$ 1.5 and
3.0, respectively. This leads to slightly different results on the
initial-final mass relations obtained with the TiO and SED calibration results
particularly at solar metallicity for which the difference between
$\alpha_\mathrm{TiO}$ and $\alpha_\mathrm{SED}$ is significant. 

The impact of the mixing length on the final mass is minor compared to that of
the overshooting parameter. A larger $f_\mathrm{ov}$ leads to substantially
higher luminosities for a given initial mass and the corresponding mass-loss
rates are higher throughout the whole evolutionary stages.  For example, the
Ledoux models with $M_\mathrm{init}=25~ M_\odot$ and $\alpha = 2.0$ at solar
metallicity have final masses of 16.1 and 12.9 for $f_\mathrm{ov} = $ 0.05 and
0.30, respectively.  The relations of the initial mass - the final helium core
and hydrogen envelope masses are also significantly affected by the
overshooting accordingly. A larger $f_\mathrm{ov}$ results in  a smaller final
mass, a larger helium core mass, and a smaller hydrogen-envelope mass.

Note that the metallicity dependence of the final mass for a given initial set
of physical parameters appears stronger in the Schwarzschild models than in the
Ledoux models. In the Ledoux models the final mass for a given initial mass
does not change significantly with metallicity, while in the Schwarzschild
models the final mass becomes much higher for a lower metallicity.   For
example, with the Schwarzschild criterion and $f_\mathrm{ov} = 0.15$,  a star
with $M_\mathrm{init}=35M_{\sun}$ is predicted to have $M_\mathrm{final} \approx
30M_{\sun}$ at SMC metallicity and  $M_\mathrm{final} \approx 18~M_\sun$ at
solar metallicity.  With the Ledoux criterion, the corresponding final masses
are $M_\mathrm{final} \approx 18~M_\odot$ and  $16 M_\odot$, respectively (see
Figures~\ref{SMCrad} and \ref{MKrad}).  This can be explained as the following.
In the Dutch scheme for mass loss of the MESA code, the mass-loss rate
prescription for RSG stars by \citet{deJager1988} does not consider a
metallicity dependence, while the mass-loss rate for hot stars is given by a
function of metallicity (i.e., $\dot{M} \propto Z^{0.85}$) as suggested by
\citet{Vink2001}. With the Ledoux criterion, stars quickly crosses the
Hertzsprung gap once hydrogen is exhausted in the core and spend the rest of
the lifetime on the Hayashi line as RSGs for the metallicities considered in
our study (Figure~\ref{fig:Teff_Yc}).  Given that mass-loss is usually more
important during the RSG phase than on the main sequence, the final masses of
the Ledoux models do not sensitively depend on the metallicity.  With the
Schwarzschild criterion,  metal-rich models ($Z \ge Z_\odot$ generally behave
like the Ledoux models although the blue loop is found for some initial masses.
However, at sub-solar metallicity, the Schwarzschild models tend to spend most
of the post-main sequence lifetime as a BSG as shown in
Figure~\ref{fig:Teff_Yc}.  The  mass-loss rates of such BSGs with
$T_\mathrm{eff} < 20000~$K are higher than those of the corresponding main
sequence stars~\citep{Vink2001},  and can play a major role for
the final mass if a star spends most of the post-main sequence phase as a BSG.
However, the BSG mass-loss rates are much lower than those of RSGs for a given
luminosity and decrease with decreasing metallicity, according to the
prescription of~\citet{Vink2001}. This can explain the reason why the
Schwarzschild models of SMC and LMC metallicities have much higher final masses
than the corresponding Ledoux models. These different predictions of
the Schwarzschild and Ledoux models can be tested with observations, in principle, 
in particular by looking at the BSG/RSG number ratio as a function of metallicity
as mentioned in Section~\ref{sec:effects} above.  

It is also important to note that all of our Ledoux models have
$M_\mathrm{H-env} < 10~M_\odot$ regardless of metallicity, while the
Schwarzschild models at sub-solor metallicities can have $ 10~M_\odot <
M_\mathrm{H-env} < 20~M_\odot$ for $M_\mathrm{init} \gtrsim 15~M_\odot$. Given
that the hydrogen envelope mass is strongly correlated with the plateau
duration and luminosity of a SN IIP,  this prediction can be tested if a good
statistics of SNe IIP from metal poor environments can be provided in future SN
surveys.  For example, \citet{Utrobin2009} argues for a very massive hydrogen
envelope mass ($M_\mathrm{H-env} \approx 14~M_\odot$) in the SN IIP 2004et,
which cannot be explained by our Ledoux models.  However, one of the reasons
for the relatively small hydrogen envelope masses with the Ledoux criterion is
that the RSG mass-loss rates are not assumed to decrease with decreasing
metallicity. No strong evidence for the metallicity dependence of the RSG wind mass-loss rate is
found so far, which is still a matter of debate~\citep[see][for a review]{vanLoon2006}. 
Another caveat here is that 
no SN IIP
progenitors with $M_\mathrm{init} \gtrsim 16~M_\odot$ have been robustly identified yet
~\citep{Smartt2009, Smartt2015}. This might imply that more massive stars
are likely to collapse to a BH, in which case bright SNe IIP from massive
progenitors with $M_\mathrm{H-env} > 10 M_\odot$ would be rare
even if the prediction of the Schwarzschild models was correct.

\subsection{Radius}\label{sec:radius}

\begin{figure}
\centering
\includegraphics[width=0.9\columnwidth]{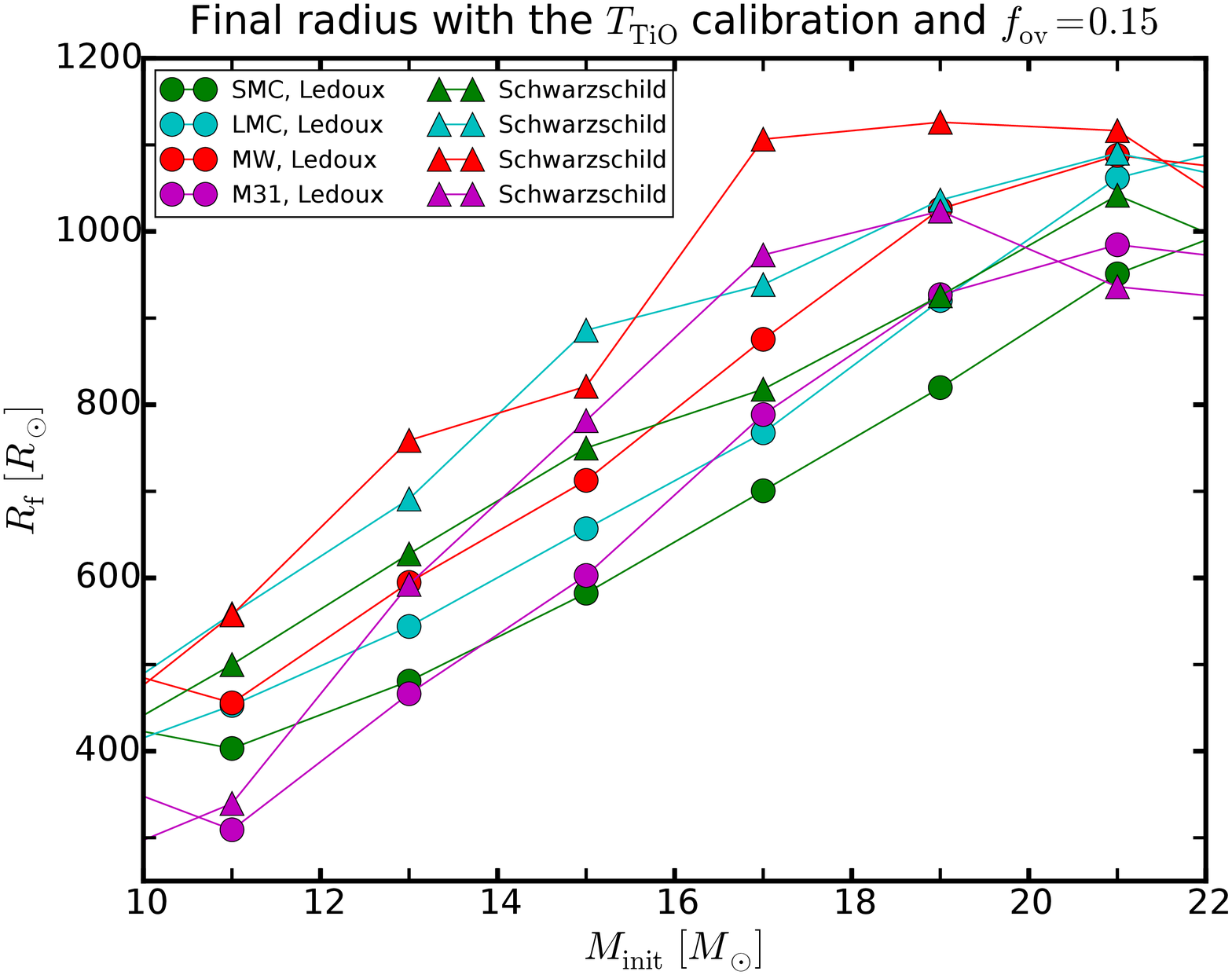}
\includegraphics[width=0.9\columnwidth]{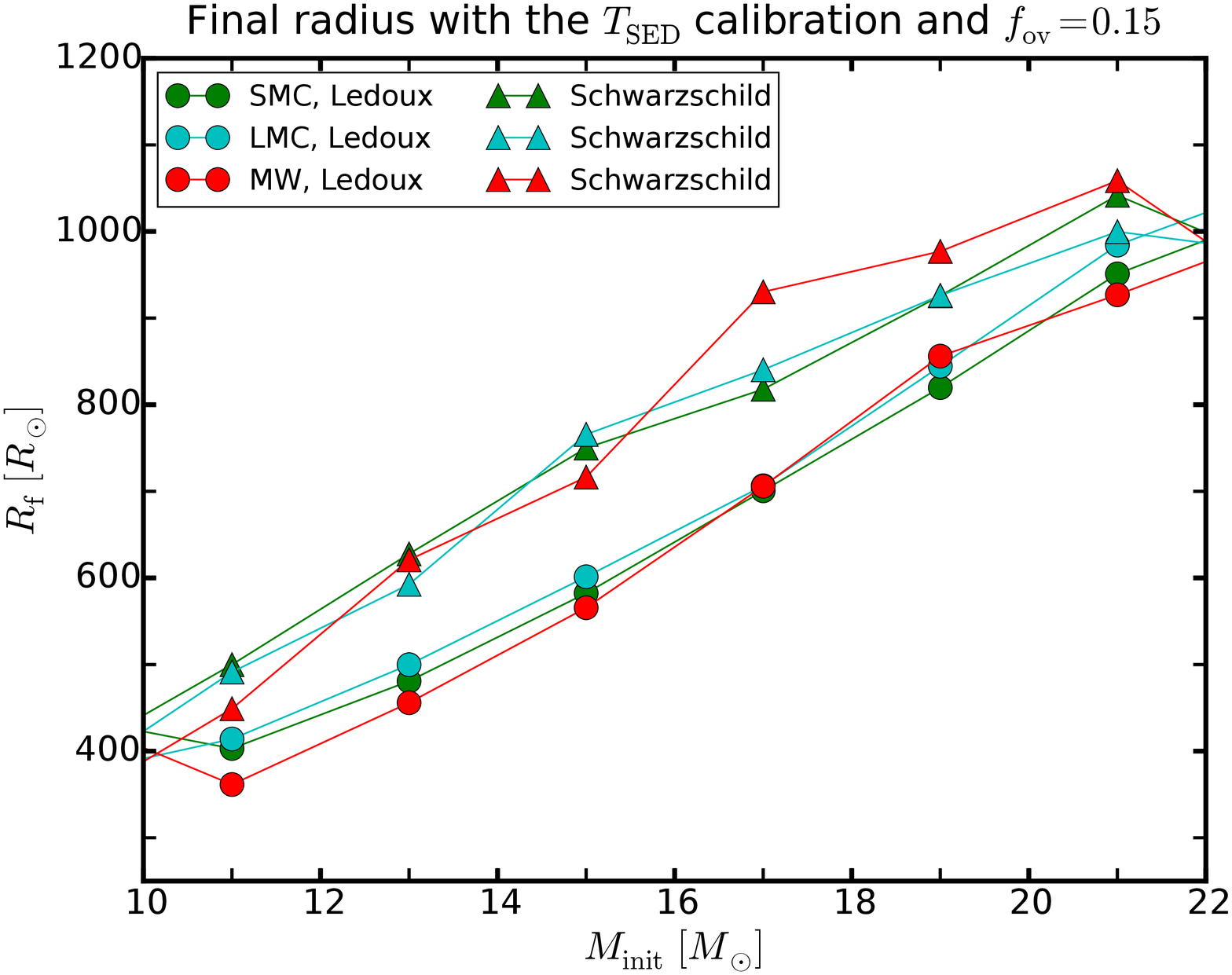}
\caption{Predicted final radii of Type IIP progenitors 
with our calibrated mixing lengths and $f_\mathrm{ov} = 0.15$. 
The TiO and SED calibration results
are given in the upper and lower panels, respectively.  
The filled circles and triangles denote the predictions with the Schwarzschild and 
Ledoux models, respectively. The different metallicities
are indicated by different colors: green ($Z=0.004$; SMC), sky blue ($Z=0.007$; LMC), 
red ($Z=0.02$; Milky Way), and purple ($Z=0.04$; M31).  
}
\label{Rfinal}
\end{figure}

One of the best ways to infer the radii of SN IIP progenitors is to compare the
theoretically predicted light curves and colors of SNe IIP with
observations~\citep[e.g.,][]{Nakar2010, Rabinak2011, Morozova2016,
Shussman2016}.   \citet{Dessart2013}, for example, concluded that supernova
models with $R \lesssim 500~R_\odot$ can best explain the U-band evolution of
typical SNe IIP and that a larger size  leads to too blue colors compared to
observations.  \citet{Gonzalez2015} measured the rise-times of light
curves for a large sample of observed SNe IIP and found that the average rise
time is $7.5\pm0.3$ d. By comparing this result with theoretical predictions,
they concluded that $R \lesssim 400~R_\odot$ is necessary to explain this short
rise-time. Shock breakout and early time observations of the SNe IIP  KSN2011a
and KSN2011d also imply relatively small radii of their progenitors (i.e.,
$\sim 280 R_\odot$ and $\sim 490 R_\odot$, respectively,
\citealt{Garnavich2016}).  Some other studies suggested that 
larger radii than $\sim 500 R_\odot$ can still be consistent with
observations~\citep[e.g.,][]{Utrobin2009, Valenti2014, Bose2015, Morozova2016}.  The caveat
in these conclusions is that the early time evolution of SNe IIP can be
strongly affected by the presence of dense circumstellar
material~\citep[e.g.,][]{Gonzalez2015, Garnavich2016, Morozova2017, Moriya2017,
Dessart2017}. 

Our mixing length calibration allows us to predict SN IIP progenitor sizes that
can be most consistent with the observed properties of RSGs.  In
Figure~\ref{Rfinal}, we present the predicted final radii of SN IIP progenitors
based on our TiO and SED calibration results, for the initial mass range of $11 -
21~M_\odot$ (see also Figures~\ref{SMCrad}, \ref{LMCrad}, \ref{MKrad}, and
\ref{M31rad}).  Some of our models with $M_\mathrm{init} = 9.0~M_\odot$ have
been followed only up to the core helium exhaustion, and are not suitable for
the prediction of the final stage. For $M_\mathrm{init} > 21~M_\odot$, the
effect of MLT++ that might lead to significant underestimates of RSG radii is
too strong (see Section~\ref{sec:comp}).  Observations also imply that SNe IIP
from such massive progenitors are rare. From the figure, we make the following
remarks.  

Firstly, there is no clear metallicity dependence of the progenitor radius.
The TiO calibration gives a systematically larger radius for a given initial
mass as the metallicity increases from $Z=0.004$ (SMC) to $Z=0.02$ (solar) but
this trend is not extended to $Z=0.04$ (M31) for which the predicted radii are
smaller than those of the solar metallicity for a given convection criterion.
Note also that the M31 models have even smaller radii than the SMC models for
$M_\mathrm{init} \lesssim 15 M_\odot$.  With the SED calibration, the scatter
due to metallicity is much smaller than in the TiO case. For the initial mass
range of  $10 \lesssim M_\mathrm{init} \lesssim 16~M_\odot$ where the
majority of SNe II are expected~\citep{Smartt2009}, the final radius ranges
from 400~$R_\odot$ to 600~$R_\odot$ with the Ledoux criterion and from from
400~$R_\odot$ to 800~$R_\odot$ with the Schwarzschild criterion, regardless of
metallicity.

Secondly, the relatively small radii of $R \lesssim 500~R_\odot$ for SN IIP
progenitors suggested by \citet{Dessart2013} and \citet{Gonzalez2015} agree
best with the predictions given by the Ledoux models with the SED calibration.
However, the very small radius of $R \approx 280~R_\odot$ inferred for the
progenitor of the SN IIP KSN2011a~\citep{Garnavich2016} is found only with
$M_\mathrm{init} \le 11 M_{\odot}$ at M31 metallicity. In general, $R \gtrsim
400~R_\odot$ is predicted within our considered parameter space.

\section{Conclusions}\label{sec:conclusions}

We have presented RGS models with the Schwarzschild and Ledoux criteria using the
MESA code, and calibrated the mixing length parameter at SMC, LMC, Milky Way,
and M31 metallicities by comparing the effective temperatures given by our RSG
models with the empirical RSG temperatures inferred from the TiO band and
SED~(Section~\ref{sec:comp}).  We also discussed its implications for SN IIP
progenitors~(Section~\ref{sec:final}). 

The main conclusion of this study is that the mixing length in RSGs depends on
metallicity. For both cases of TiO and SED temperatures, we find that the
mixing length is an increasing function of metallicity (Table~\ref{tab1} and
Figure~\ref{mixing}).  Our finding probably indicates that the efficiency of
the convective energy transport in RSGs becomes higher for higher metallicity.
This result is in qualitatively accordance with the recent finding of the
correlation between mixing length and metallicity in low-mass red giant stars
by \citet{Tayar2017}, implying that this correlation is a universal feature in
post-main sequence stars for both low and high masses.  Currently, there exists
no theory that can explain this tendency and future studies should address this
important issue.  We should also investigate  if this correlation 
can be extended to a metallicity beyond our considered range.  

For our study, we have investigated the code dependencies of RSG models, and
found that the Hayashi lines predicted from different numerical methods
including MESA, BEC, and TWIN codes agree remarkably well, and therefore our
calibrated mixing length values may be adopted in other stellar evolution codes
that solve the same set of stellar structure equations as in these codes~(See
Section~\ref{sec:effects}).  However, the models by the Geneva group give
significantly higher RSG temperatures for a given mixing length parameter,
which calls for a future investigation on the impact of
numerical schemes employed in different codes.

The final structures of RSGs given by our calibrated mixing length can provide
useful predictions on the properties of SN IIP supernova
progenitors~(Section~\ref{sec:final}). In particular, the final radii are
expected to be about $400~R_\odot - 800~R_\odot$ for the initial mass range of
$10~M_\odot \lesssim M_\mathrm{init} \lesssim 16~M_\odot$ which is typical for
SN IIP progenitors~(Section~\ref{sec:radius}).  Our result also implies that
the radii of SN IIP progenitors for a given initial mass do not depend on
metallicity.  

Another important finding in this study (although it is not directly related to
the mixing length) is that, for $M_\mathrm{init} \gtrsim 15~M_\odot$, the
hydrogen envelope masses of SN IIP progenitors at SMC and LMC metallicities can
be much higher with the Schwarzschild criterion ($M_\mathrm{H-env} \simeq 10 -
20$) than with the Ledoux criterion and slow semi-convection ($M_\mathrm{H-env}
< 10 M_\odot$).  In the latter case the hydrogen envelope mass does not appear to strongly
depend on metallicity (Section~\ref{sec:masses}). This could be tested in
principle with a sufficiently large sample of SNe IIP from metal poor
environments~\citep[cf.][]{Anderson2016}.

\acknowledgments

We are grateful to Georges Meynet, who refereed this paper, 
for many constructive comments that helped us improve the paper. 
This work was supported by the Basic Science Research program
(NRF-2016R1D1A1A09918398) through the National Research Foundation of Korea
(NRF) and by the Korea Astronomy and Space Science Institute under the R\&D
program (Project No. 3348- 20160002) supervised by the Ministry of Science and
ICT.  SCY acknowledges the support by the Monash Center for Astrophysics via
the distinguished visitor program. 

\appendix
\section{Appendix Material}

In the figures, we present the final radius, final total mass, and final mass
of the hydrogen-rich envelope as a function of the initial mass, which are
obtained by interpolating the results of our model sequences at the calibrated
mixing length values. Here we excluded models with $M_\mathrm{H-env} < 0.5 M_\odot$.

\begin{figure*}
\centering
\includegraphics[width=0.9\textwidth]{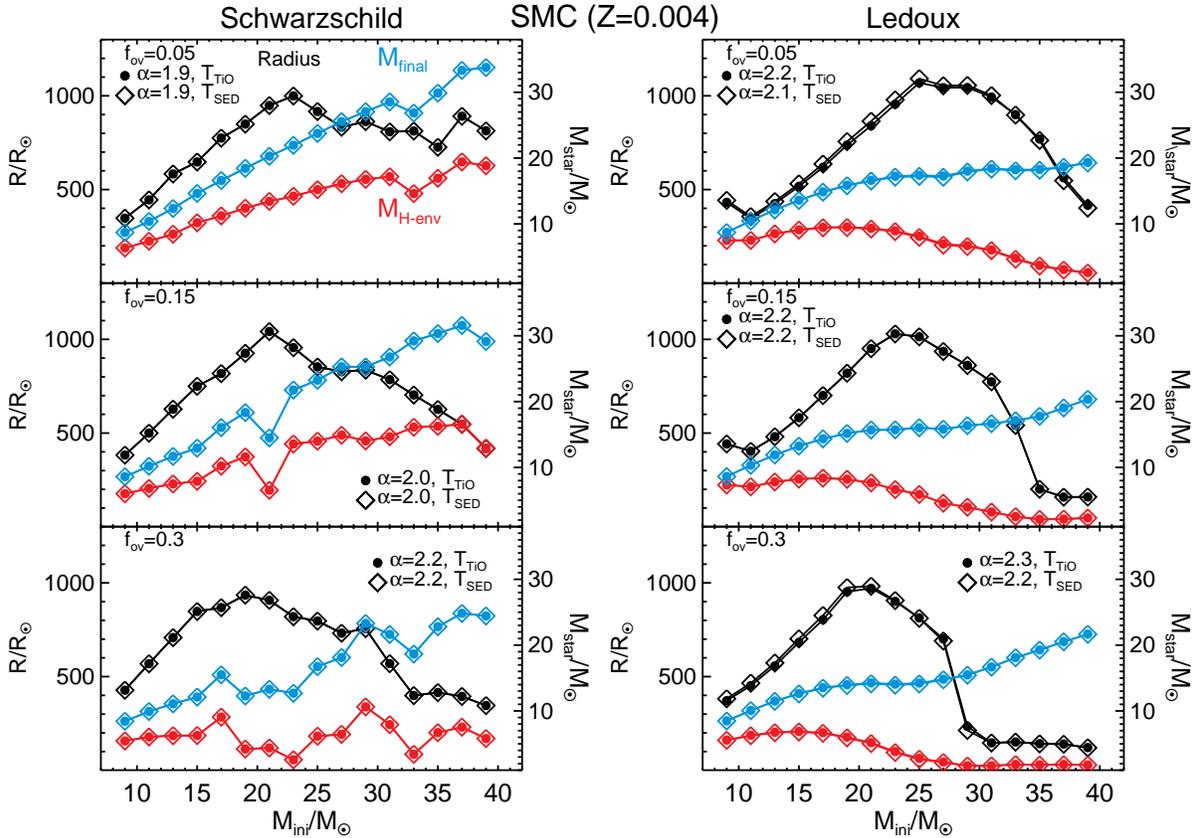}
\caption{Final radius, final total mass,  and final  hydrogen-rich envelope mass
as a function of the initial mass
for the SMC metallicity (Z=0.004) predicted from our
mixing length calibration with TiO (filled circle) and SED (open diamond) temperatures.  
The left and right panels present the results with the Schwarzschild and Ledoux models, respectively. 
The final radii are plotted as black color and their size indicated on the left axis.
The final total mass and final hydrogen-rich envelope mass are indicated by blue and red colors, respectively,
and their masses are indicated on the right axis.
The results of three different overshooting parameters ($f_\mathrm{ov}=0.05, 0.15,$ and $0.3$) are 
plotted from the top to bottom panels. In each panel, 
the calibrated mixing length values by TiO and SED temperatures for the given
metallicity and overshooting parameter are indicated by different symbols.
Here we excluded models with $M_\mathrm{H-env} < 0.5 M_\odot$. 
}
\label{SMCrad}
\end{figure*}

\begin{figure*}
\centering
\includegraphics[width=0.9\textwidth]{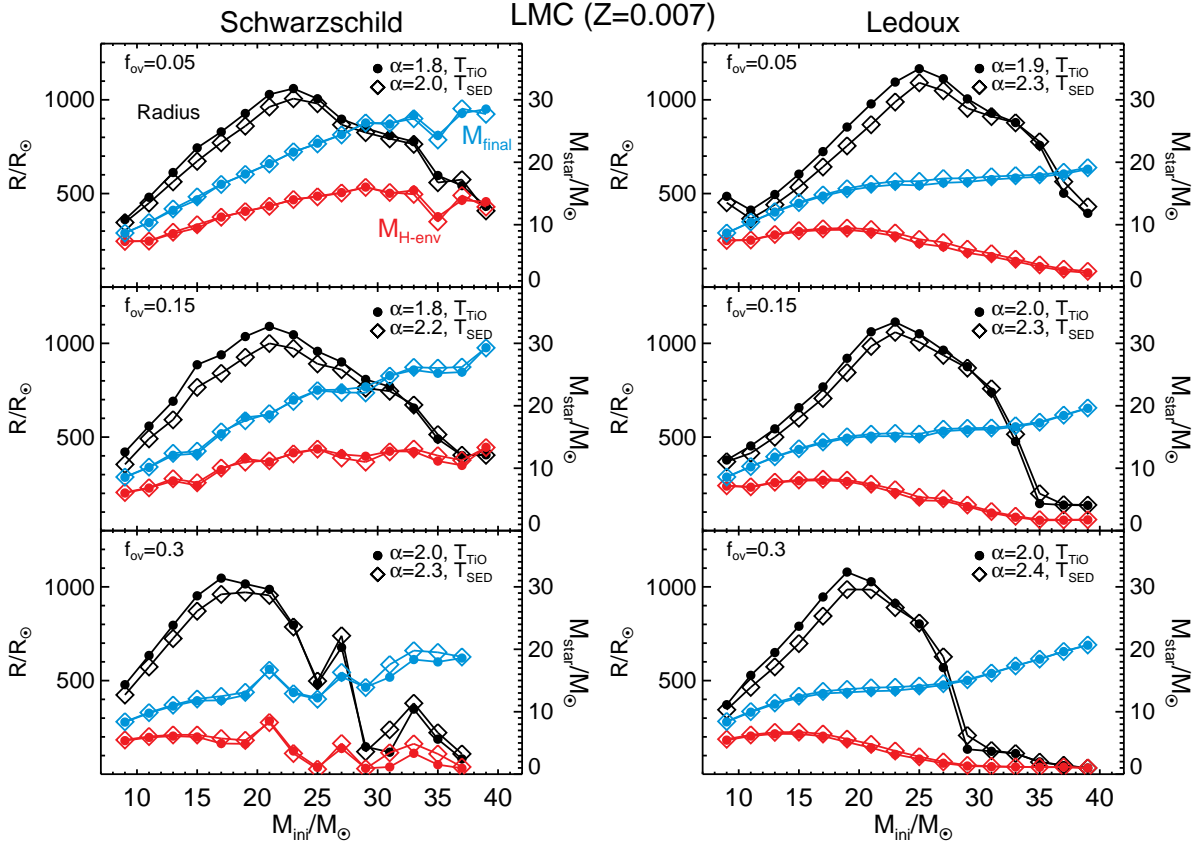}
\caption{Same with Figure~\ref{SMCrad} but for LMC metallicity (Z=0.007). 
}
\label{LMCrad}
\end{figure*}

\begin{figure*}
\centering
\includegraphics[width=0.9\textwidth]{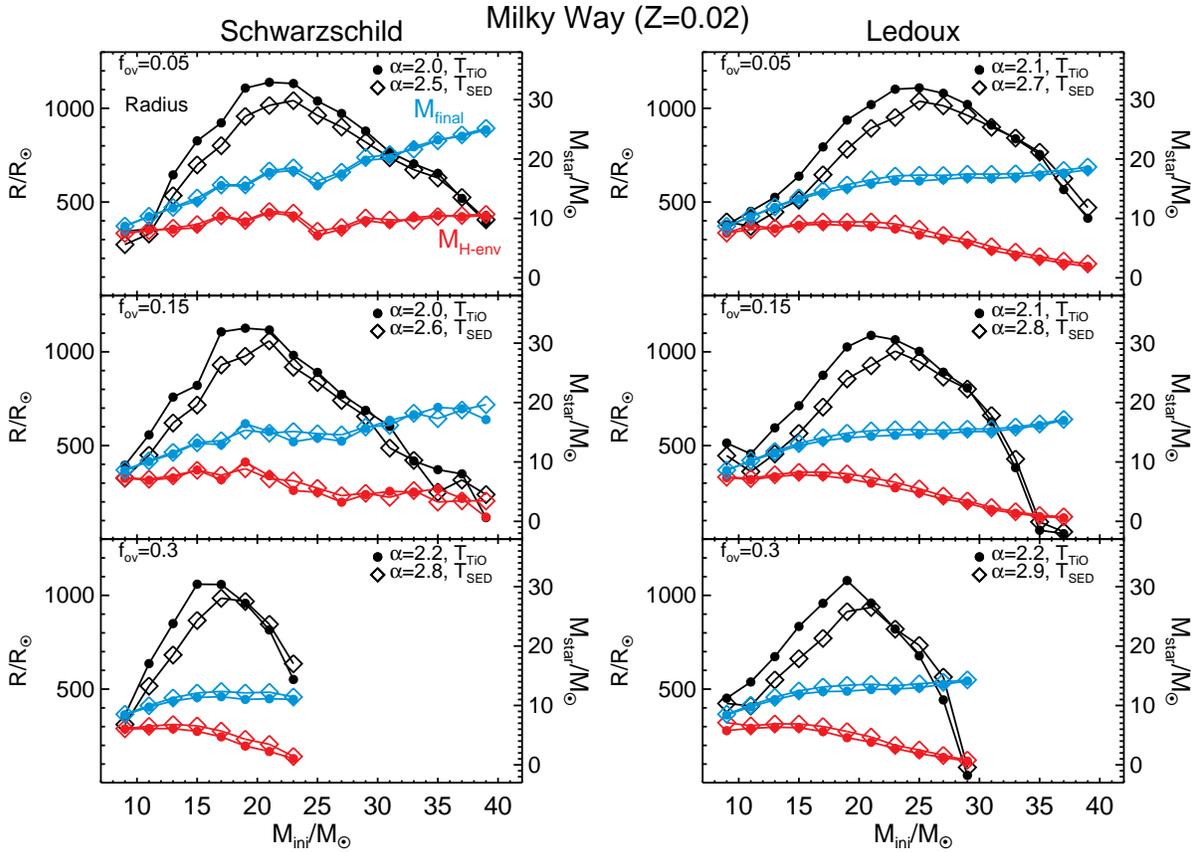}
\caption{Same with Figure~\ref{SMCrad} but for solar metallicity (Z=0.02).
}
\label{MKrad}
\end{figure*}

\begin{figure*}
\centering
\includegraphics[width=0.9\textwidth]{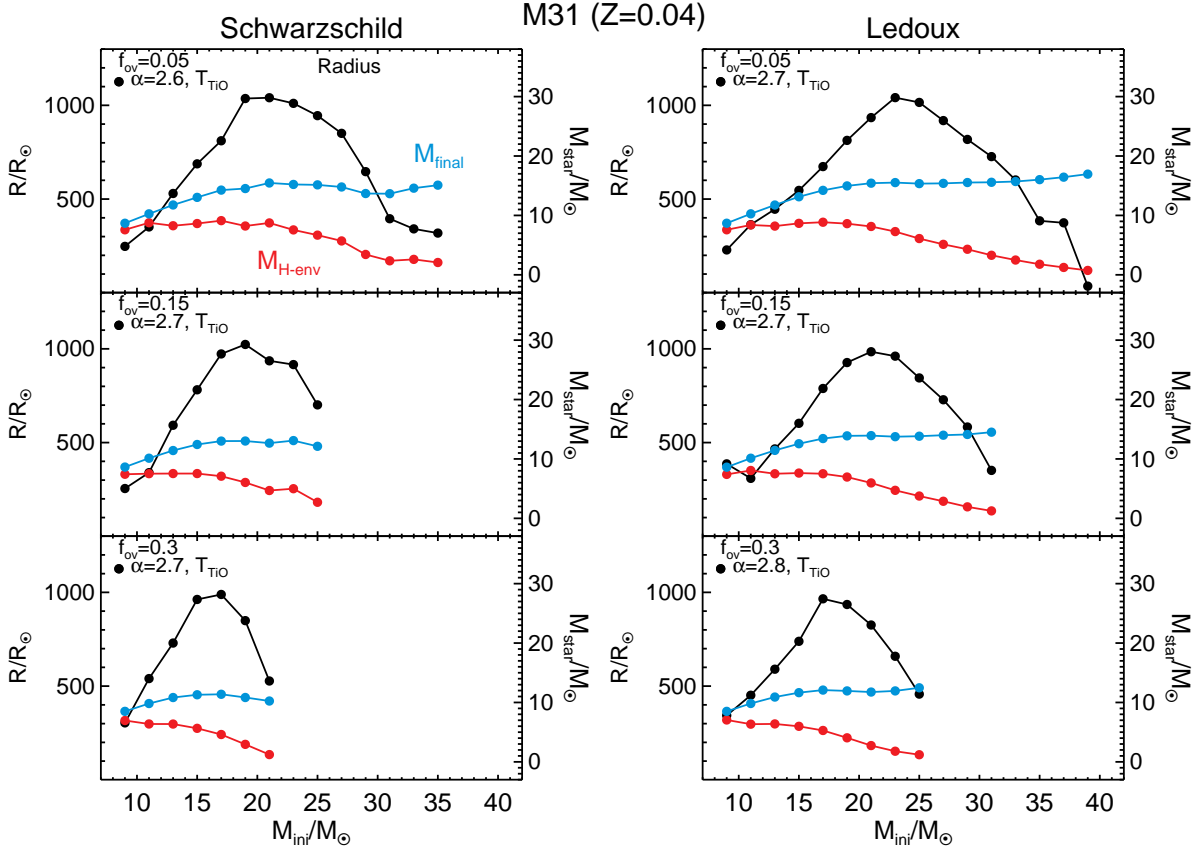}
\caption{Same with Figure~\ref{SMCrad} but for M31 metallicity (Z=0.04).
}
\label{M31rad}
\end{figure*}

\bibliographystyle{aasjournal}
\bibliography{reference}

\begin{thebibliography}{}
\expandafter\ifx\csname natexlab\endcsname\relax\def\natexlab#1{#1}\fi

\bibitem[{{Alongi} {et~al.}(1991){Alongi}, {Bertelli}, {Bressan}, \&
  {Chiosi}}]{Alongi1991}
{Alongi}, M., {Bertelli}, G., {Bressan}, A., \& {Chiosi}, C. 1991, \aap, 244,
  95

\bibitem[{{Anderson} {et~al.}(2016){Anderson}, {Guti{\'e}rrez}, {Dessart},
  {Hamuy}, {Galbany}, {Morrell}, {Stritzinger}, {Phillips}, {Folatelli},
  {Boffin}, {de Jaeger}, {Kuncarayakti}, \& {Prieto}}]{Anderson2016}
{Anderson}, J.~P., {Guti{\'e}rrez}, C.~P., {Dessart}, L., {et~al.} 2016, \aap,
  589, A110

\bibitem[{{Asplund} {et~al.}(2005){Asplund}, {Grevesse}, \&
  {Sauval}}]{Asplund2005}
{Asplund}, M., {Grevesse}, N., \& {Sauval}, A.~J. 2005, in Astronomical Society
  of the Pacific Conference Series, Vol. 336, Cosmic Abundances as Records of
  Stellar Evolution and Nucleosynthesis, ed. T.~G. {Barnes}, III \& F.~N.
  {Bash}, 25

\bibitem[{{Asplund} {et~al.}(2009){Asplund}, {Grevesse}, {Sauval}, \&
  {Scott}}]{Asplund2009}
{Asplund}, M., {Grevesse}, N., {Sauval}, A.~J., \& {Scott}, P. 2009, \araa, 47,
  481

\bibitem[{{Bono} {et~al.}(2000){Bono}, {Caputo}, {Cassisi}, {Marconi},
  {Piersanti}, \& {Tornamb{\`e}}}]{Bono2000}
{Bono}, G., {Caputo}, F., {Cassisi}, S., {et~al.} 2000, \apj, 543, 955

\bibitem[{{Bose} {et~al.}(2015){Bose}, {Valenti}, {Misra}, {Pumo}, {Zampieri},
  {Sand}, {Kumar}, {Pastorello}, {Sutaria}, {Maccarone}, {Kumar}, {Graham},
  {Howell}, {Ochner}, {Chandola}, \& {Pandey}}]{Bose2015}
{Bose}, S., {Valenti}, S., {Misra}, K., {et~al.} 2015, \mnras, 450, 2373

\bibitem[{{Brott} {et~al.}(2011){Brott}, {de Mink}, {Cantiello}, {Langer}, {de
  Koter}, {Evans}, {Hunter}, {Trundle}, \& {Vink}}]{Brott2011}
{Brott}, I., {de Mink}, S.~E., {Cantiello}, M., {et~al.} 2011, \aap, 530, A115

\bibitem[{{Chiavassa} {et~al.}(2011){Chiavassa}, {Freytag}, {Masseron}, \&
  {Plez}}]{Chiavassa2011}
{Chiavassa}, A., {Freytag}, B., {Masseron}, T., \& {Plez}, B. 2011, \aap, 535,
  A22

\bibitem[{{Chieffi} {et~al.}(1995){Chieffi}, {Straniero}, \&
  {Salaris}}]{Chieffi1995}
{Chieffi}, A., {Straniero}, O., \& {Salaris}, M. 1995, \apjl, 445, L39

\bibitem[{{Cunha} {et~al.}(2006){Cunha}, {Hubeny}, \& {Lanz}}]{Cunha2006}
{Cunha}, K., {Hubeny}, I., \& {Lanz}, T. 2006, \apjl, 647, L143

\bibitem[{{Davies} {et~al.}(2015){Davies}, {Kudritzki}, {Gazak}, {Plez},
  {Bergemann}, {Evans}, \& {Patrick}}]{Davies2015}
{Davies}, B., {Kudritzki}, R.-P., {Gazak}, Z., {et~al.} 2015, \apj, 806, 21

\bibitem[{{Davies} {et~al.}(2013){Davies}, {Kudritzki}, {Plez}, {Trager},
  {Lan{\c c}on}, {Gazak}, {Bergemann}, {Evans}, \& {Chiavassa}}]{Davies2013}
{Davies}, B., {Kudritzki}, R.-P., {Plez}, B., {et~al.} 2013, \apj, 767, 3

\bibitem[{{de Jager} {et~al.}(1988){de Jager}, {Nieuwenhuijzen}, \& {van der
  Hucht}}]{deJager1988}
{de Jager}, C., {Nieuwenhuijzen}, H., \& {van der Hucht}, K.~A. 1988, \aaps,
  72, 259

\bibitem[{{Dessart} {et~al.}(2017){Dessart}, {Hillier}, \&
  {Audit}}]{Dessart2017}
{Dessart}, L., {Hillier}, D.~J., \& {Audit}, E. 2017, ArXiv e-prints,
  arXiv:1704.01697

\bibitem[{{Dessart} {et~al.}(2013){Dessart}, {Hillier}, {Waldman}, \&
  {Livne}}]{Dessart2013}
{Dessart}, L., {Hillier}, D.~J., {Waldman}, R., \& {Livne}, E. 2013, \mnras,
  433, 1745

\bibitem[{{Drout} {et~al.}(2009){Drout}, {Massey}, {Meynet}, {Tokarz}, \&
  {Caldwell}}]{Drout2009}
{Drout}, M.~R., {Massey}, P., {Meynet}, G., {Tokarz}, S., \& {Caldwell}, N.
  2009, \apj, 703, 441

\bibitem[{{Eggenberger} {et~al.}(2002){Eggenberger}, {Meynet}, \&
  {Maeder}}]{Eggenberger2002}
{Eggenberger}, P., {Meynet}, G., \& {Maeder}, A. 2002, \aap, 386, 576

\bibitem[{{Eggleton}(1971)}]{Eggleton1971}
{Eggleton}, P.~P. 1971, \mnras, 151, 351

\bibitem[{{Eggleton} \& {Kiseleva-Eggleton}(2002)}]{Eggleton2002}
{Eggleton}, P.~P., \& {Kiseleva-Eggleton}, L. 2002, \apj, 575, 461

\bibitem[{{Ekstr{\"o}m} {et~al.}(2012){Ekstr{\"o}m}, {Georgy}, {Eggenberger},
  {Meynet}, {Mowlavi}, {Wyttenbach}, {Granada}, {Decressin}, {Hirschi},
  {Frischknecht}, {Charbonnel}, \& {Maeder}}]{Ekstrom2012}
{Ekstr{\"o}m}, S., {Georgy}, C., {Eggenberger}, P., {et~al.} 2012, \aap, 537,
  A146

\bibitem[{{El Eid}(1995)}]{El1995}
{El Eid}, M.~F. 1995, \mnras, 275, 983

\bibitem[{{Eldridge} {et~al.}(2013){Eldridge}, {Fraser}, {Smartt}, {Maund}, \&
  {Crockett}}]{Eldridge2013}
{Eldridge}, J.~J., {Fraser}, M., {Smartt}, S.~J., {Maund}, J.~R., \&
  {Crockett}, R.~M. 2013, \mnras, 436, 774

\bibitem[{{Elias} {et~al.}(1985){Elias}, {Frogel}, \& {Humphreys}}]{Elias1985}
{Elias}, J.~H., {Frogel}, J.~A., \& {Humphreys}, R.~M. 1985, \apjs, 57, 91

\bibitem[{{Gabriel} {et~al.}(2014){Gabriel}, {Noels}, {Montalb{\'a}n}, \&
  {Miglio}}]{Gabriel2014}
{Gabriel}, M., {Noels}, A., {Montalb{\'a}n}, J., \& {Miglio}, A. 2014, \aap,
  569, A63

\bibitem[{{Garnavich} {et~al.}(2016){Garnavich}, {Tucker}, {Rest}, {Shaya},
  {Olling}, {Kasen}, \& {Villar}}]{Garnavich2016}
{Garnavich}, P.~M., {Tucker}, B.~E., {Rest}, A., {et~al.} 2016, \apj, 820, 23

\bibitem[{{Gazak} {et~al.}(2014){Gazak}, {Davies}, {Bastian}, {Kudritzki},
  {Bergemann}, {Plez}, {Evans}, {Patrick}, {Bresolin}, \&
  {Schinnerer}}]{Gazak2014}
{Gazak}, J.~Z., {Davies}, B., {Bastian}, N., {et~al.} 2014, \apj, 787, 142

\bibitem[{{Gazak} {et~al.}(2015){Gazak}, {Kudritzki}, {Evans}, {Patrick},
  {Davies}, {Bergemann}, {Plez}, {Bresolin}, {Bender}, {Wegner}, {Bonanos}, \&
  {Williams}}]{Gazak2015}
{Gazak}, J.~Z., {Kudritzki}, R., {Evans}, C., {et~al.} 2015, \apj, 805, 182

\bibitem[{{Georgy}(2012)}]{Georgy2012}
{Georgy}, C. 2012, \aap, 538, L8

\bibitem[{{Georgy} {et~al.}(2013){Georgy}, {Ekstr{\"o}m}, {Eggenberger},
  {Meynet}, {Haemmerl{\'e}}, {Maeder}, {Granada}, {Groh}, {Hirschi}, {Mowlavi},
  {Yusof}, {Charbonnel}, {Decressin}, \& {Barblan}}]{Georgy2013}
{Georgy}, C., {Ekstr{\"o}m}, S., {Eggenberger}, P., {et~al.} 2013, \aap, 558,
  A103

\bibitem[{{Gonz{\'a}lez-Gait{\'a}n} {et~al.}(2015){Gonz{\'a}lez-Gait{\'a}n},
  {Tominaga}, {Molina}, {Galbany}, {Bufano}, {Anderson}, {Gutierrez},
  {F{\"o}rster}, {Pignata}, {Bersten}, {Howell}, {Sullivan}, {Carlberg}, {de
  Jaeger}, {Hamuy}, {Baklanov}, \& {Blinnikov}}]{Gonzalez2015}
{Gonz{\'a}lez-Gait{\'a}n}, S., {Tominaga}, N., {Molina}, J., {et~al.} 2015,
  \mnras, 451, 2212

\bibitem[{{Gordon} {et~al.}(2016){Gordon}, {Humphreys}, \&
  {Jones}}]{Gordon2016}
{Gordon}, M.~S., {Humphreys}, R.~M., \& {Jones}, T.~J. 2016, \apj, 825, 50

\bibitem[{{Grevesse} \& {Sauval}(1998)}]{Grevesse1998}
{Grevesse}, N., \& {Sauval}, A.~J. 1998, \ssr, 85, 161

\bibitem[{{Hayashi} \& {Hoshi}(1961)}]{Hayashi1961}
{Hayashi}, C., \& {Hoshi}, R. 1961, \pasj, 13, 442

\bibitem[{{Heger} {et~al.}(2000){Heger}, {Langer}, \& {Woosley}}]{Heger2000}
{Heger}, A., {Langer}, N., \& {Woosley}, S.~E. 2000, \apj, 528, 368

\bibitem[{{Izzard} \& {Glebbeek}(2006)}]{Izzard2006}
{Izzard}, R.~G., \& {Glebbeek}, E. 2006, \na, 12, 161

\bibitem[{{Jones} {et~al.}(2015){Jones}, {Hirschi}, {Pignatari}, {Heger},
  {Georgy}, {Nishimura}, {Fryer}, \& {Herwig}}]{Jones2015}
{Jones}, S., {Hirschi}, R., {Pignatari}, M., {et~al.} 2015, \mnras, 447, 3115

\bibitem[{{Kippenhahn} \& {Weigert}(1990)}]{Kipp1990}
{Kippenhahn}, R., \& {Weigert}, A. 1990, {Stellar Structure and Evolution}, 192

\bibitem[{{Langer}(2012)}]{Langer2012}
{Langer}, N. 2012, \araa, 50, 107

\bibitem[{{Langer} \& {Maeder}(1995)}]{Langer1995}
{Langer}, N., \& {Maeder}, A. 1995, \aap, 295, 685

\bibitem[{{Levesque} {et~al.}(2005){Levesque}, {Massey}, {Olsen}, {Plez},
  {Josselin}, {Maeder}, \& {Meynet}}]{Levesque2005}
{Levesque}, E.~M., {Massey}, P., {Olsen}, K.~A.~G., {et~al.} 2005, \apj, 628,
  973

\bibitem[{{Levesque} {et~al.}(2006){Levesque}, {Massey}, {Olsen}, {Plez},
  {Meynet}, \& {Maeder}}]{Levesque2006}
---. 2006, \apj, 645, 1102

\bibitem[{{Maeder} \& {Meynet}(2000)}]{Maeder2000}
{Maeder}, A., \& {Meynet}, G. 2000, \araa, 38, 143

\bibitem[{{Magic} {et~al.}(2015){Magic}, {Weiss}, \& {Asplund}}]{Magic2015}
{Magic}, Z., {Weiss}, A., \& {Asplund}, M. 2015, \aap, 573, A89

\bibitem[{{Martins} \& {Palacios}(2013)}]{Martins2013}
{Martins}, F., \& {Palacios}, A. 2013, \aap, 560, A16

\bibitem[{{Massey} \& {Evans}(2016)}]{Massey2016}
{Massey}, P., \& {Evans}, K.~A. 2016, \apj, 826, 224

\bibitem[{{Massey} {et~al.}(2009){Massey}, {Silva}, {Levesque}, {Plez},
  {Olsen}, {Clayton}, {Meynet}, \& {Maeder}}]{Massey2009}
{Massey}, P., {Silva}, D.~R., {Levesque}, E.~M., {et~al.} 2009, \apj, 703, 420

\bibitem[{{Meynet} \& {Maeder}(1997)}]{Meynet1997}
{Meynet}, G., \& {Maeder}, A. 1997, \aap, 321, 465

\bibitem[{{Meynet} {et~al.}(2015){Meynet}, {Chomienne}, {Ekstr{\"o}m},
  {Georgy}, {Granada}, {Groh}, {Maeder}, {Eggenberger}, {Levesque}, \&
  {Massey}}]{Meynet2015}
{Meynet}, G., {Chomienne}, V., {Ekstr{\"o}m}, S., {et~al.} 2015, \aap, 575, A60

\bibitem[{{Mokiem} {et~al.}(2006){Mokiem}, {de Koter}, {Evans}, {Puls},
  {Smartt}, {Crowther}, {Herrero}, {Langer}, {Lennon}, {Najarro}, {Villamariz},
  \& {Yoon}}]{Mokiem2006}
{Mokiem}, M.~R., {de Koter}, A., {Evans}, C.~J., {et~al.} 2006, \aap, 456, 1131

\bibitem[{{Moriya} {et~al.}(2017){Moriya}, {Yoon}, {Gr{\"a}fener}, \&
  {Blinnikov}}]{Moriya2017}
{Moriya}, T.~J., {Yoon}, S.-C., {Gr{\"a}fener}, G., \& {Blinnikov}, S.~I. 2017,
  \mnras, 469, L108

\bibitem[{{Morozova} {et~al.}(2016){Morozova}, {Piro}, {Renzo}, \&
  {Ott}}]{Morozova2016}
{Morozova}, V., {Piro}, A.~L., {Renzo}, M., \& {Ott}, C.~D. 2016, \apj, 829,
  109

\bibitem[{{Morozova} {et~al.}(2017){Morozova}, {Piro}, \&
  {Valenti}}]{Morozova2017}
{Morozova}, V., {Piro}, A.~L., \& {Valenti}, S. 2017, \apj, 838, 28

\bibitem[{{Nakar} \& {Sari}(2010)}]{Nakar2010}
{Nakar}, E., \& {Sari}, R. 2010, \apj, 725, 904

\bibitem[{{Patrick} {et~al.}(2015){Patrick}, {Evans}, {Davies}, {Kudritzki},
  {Gazak}, {Bergemann}, {Plez}, \& {Ferguson}}]{Patrick2015}
{Patrick}, L.~R., {Evans}, C.~J., {Davies}, B., {et~al.} 2015, \apj, 803, 14

\bibitem[{{Paxton} {et~al.}(2011){Paxton}, {Bildsten}, {Dotter}, {Herwig},
  {Lesaffre}, \& {Timmes}}]{Paxton2011}
{Paxton}, B., {Bildsten}, L., {Dotter}, A., {et~al.} 2011, \apjs, 192, 3

\bibitem[{{Paxton} {et~al.}(2013){Paxton}, {Cantiello}, {Arras}, {Bildsten},
  {Brown}, {Dotter}, {Mankovich}, {Montgomery}, {Stello}, {Timmes}, \&
  {Townsend}}]{Paxton2013}
{Paxton}, B., {Cantiello}, M., {Arras}, P., {et~al.} 2013, \apjs, 208, 4

\bibitem[{{Paxton} {et~al.}(2015){Paxton}, {Marchant}, {Schwab}, {Bauer},
  {Bildsten}, {Cantiello}, {Dessart}, {Farmer}, {Hu}, {Langer}, {Townsend},
  {Townsley}, \& {Timmes}}]{Paxton2015}
{Paxton}, B., {Marchant}, P., {Schwab}, J., {et~al.} 2015, \apjs, 220, 15

\bibitem[{{Paxton} {et~al.}(2017){Paxton}, {Schwab}, {Bauer}, {Bildsten},
  {Blinnikov}, {Duffell}, {Farmer}, {Goldberg}, {Marchant}, {Sorokina},
  {Thoul}, {Townsend}, \& {Timmes}}]{Paxton2017}
{Paxton}, B., {Schwab}, J., {Bauer}, E.~B., {et~al.} 2017, ArXiv e-prints,
  arXiv:1710.08424

\bibitem[{{Podsiadlowski} {et~al.}(1992){Podsiadlowski}, {Joss}, \&
  {Hsu}}]{Podsiadlowski1992}
{Podsiadlowski}, P., {Joss}, P.~C., \& {Hsu}, J.~J.~L. 1992, \apj, 391, 246

\bibitem[{{Rabinak} \& {Waxman}(2011)}]{Rabinak2011}
{Rabinak}, I., \& {Waxman}, E. 2011, \apj, 728, 63

\bibitem[{{Ram{\'{\i}}rez-Agudelo} {et~al.}(2013){Ram{\'{\i}}rez-Agudelo},
  {Sim{\'o}n-D{\'{\i}}az}, {Sana}, {de Koter}, {Sab{\'{\i}}n-Sanjul{\'{\i}}an},
  {de Mink}, {Dufton}, {Gr{\"a}fener}, {Evans}, {Herrero}, {Langer}, {Lennon},
  {Ma{\'{\i}}z Apell{\'a}niz}, {Markova}, {Najarro}, {Puls}, {Taylor}, \&
  {Vink}}]{Agudelo2013}
{Ram{\'{\i}}rez-Agudelo}, O.~H., {Sim{\'o}n-D{\'{\i}}az}, S., {Sana}, H.,
  {et~al.} 2013, \aap, 560, A29

\bibitem[{{Sanders} {et~al.}(2012){Sanders}, {Caldwell}, {McDowell}, \&
  {Harding}}]{Sanders2012}
{Sanders}, N.~E., {Caldwell}, N., {McDowell}, J., \& {Harding}, P. 2012, \apj,
  758, 133

\bibitem[{{Schaller} {et~al.}(1992){Schaller}, {Schaerer}, {Meynet}, \&
  {Maeder}}]{Schaller1992}
{Schaller}, G., {Schaerer}, D., {Meynet}, G., \& {Maeder}, A. 1992, \aaps, 96,
  269

\bibitem[{{Shussman} {et~al.}(2016){Shussman}, {Waldman}, \&
  {Nakar}}]{Shussman2016}
{Shussman}, T., {Waldman}, R., \& {Nakar}, E. 2016, ArXiv e-prints,
  arXiv:1610.05323

\bibitem[{{Sim{\'o}n-D{\'{\i}}az} \& {Herrero}(2014)}]{Simon2014}
{Sim{\'o}n-D{\'{\i}}az}, S., \& {Herrero}, A. 2014, \aap, 562, A135

\bibitem[{{Smartt}(2009)}]{Smartt2009}
{Smartt}, S.~J. 2009, \araa, 47, 63

\bibitem[{{Smartt}(2015)}]{Smartt2015}
---. 2015, \pasa, 32, e016

\bibitem[{{Smith}(2014)}]{Smith2014}
{Smith}, N. 2014, \araa, 52, 487

\bibitem[{{Stothers} \& {Chin}(1991)}]{Stothers1991}
{Stothers}, R.~B., \& {Chin}, C.-W. 1991, \apj, 374, 288

\bibitem[{{Stothers} \& {Chin}(1996)}]{Stothers1996}
---. 1996, \apj, 469, 166

\bibitem[{{Tayar} {et~al.}(2017){Tayar}, {Somers}, {Pinsonneault}, {Stello},
  {Mints}, {Johnson}, {Zamora}, {Garc{\'{\i}}a-Hern{\'a}ndez}, {Maraston},
  {Serenelli}, {Allende Prieto}, {Bastien}, {Basu}, {Bird}, {Cohen}, {Cunha},
  {Elsworth}, {Garc{\'{\i}}a}, {Girardi}, {Hekker}, {Holtzman}, {Huber},
  {Mathur}, {M{\'e}sz{\'a}ros}, {Mosser}, {Shetrone}, {Silva Aguirre},
  {Stassun}, {Stringfellow}, {Zasowski}, \& {Roman-Lopes}}]{Tayar2017}
{Tayar}, J., {Somers}, G., {Pinsonneault}, M.~H., {et~al.} 2017, \apj, 840, 17

\bibitem[{{Utrobin} \& {Chugai}(2009)}]{Utrobin2009}
{Utrobin}, V.~P., \& {Chugai}, N.~N. 2009, \aap, 506, 829

\bibitem[{{Valenti} {et~al.}(2014){Valenti}, {Sand}, {Pastorello}, {Graham},
  {Howell}, {Parrent}, {Tomasella}, {Ochner}, {Fraser}, {Benetti}, {Yuan},
  {Smartt}, {Maund}, {Arcavi}, {Gal-Yam}, {Inserra}, \& {Young}}]{Valenti2014}
{Valenti}, S., {Sand}, D., {Pastorello}, A., {et~al.} 2014, \mnras, 438, L101

\bibitem[{{van Loon}(2006)}]{vanLoon2006}
{van Loon}, J.~T. 2006, in Astronomical Society of the Pacific Conference
  Series, Vol. 353, Stellar Evolution at Low Metallicity: Mass Loss,
  Explosions, Cosmology, ed. H.~J.~G.~L.~M. {Lamers}, N.~{Langer}, T.~{Nugis},
  \& K.~{Annuk}, 211

\bibitem[{{Vink} {et~al.}(2001){Vink}, {de Koter}, \& {Lamers}}]{Vink2001}
{Vink}, J.~S., {de Koter}, A., \& {Lamers}, H.~J.~G.~L.~M. 2001, \aap, 369, 574

\bibitem[{{Yoon} \& {Cantiello}(2010)}]{Yoon2010}
{Yoon}, S.-C., \& {Cantiello}, M. 2010, \apjl, 717, L62

\bibitem[{{Yoon} {et~al.}(2017){Yoon}, {Dessart}, \& {Clocchiatti}}]{Yoon2017}
{Yoon}, S.-C., {Dessart}, L., \& {Clocchiatti}, A. 2017, \apj, 840, 10

\bibitem[{{Zaussinger} \& {Spruit}(2013)}]{Zaussinger2013}
{Zaussinger}, F., \& {Spruit}, H.~C. 2013, \aap, 554, A119

\end{thebibliography}
\end{document}